\documentclass[traditabstract]{aa}

\usepackage[english]{babel}
\usepackage{graphicx}
\usepackage{amsmath}
\usepackage{amssymb}
\usepackage{txfonts}
\usepackage{wasysym}
\usepackage{gensymb}

\usepackage[singlelinecheck=true,format=default,font=small,labelfont=bf]{caption}
\usepackage{natbib}
\usepackage{color}
\usepackage{multirow}
\usepackage{array}

\usepackage{upgreek}

\bibpunct{(}{)}{;}{a}{}{,} 
\defcitealias{Fritz2012}{HELGA I}
\defcitealias{Smith2012}{HELGA II}
\defcitealias{Ford2013}{HELGA III}
\defcitealias{Viaene2014}{HELGA IV}

\begin{document}

\title{The \textit{Herschel}\thanks{\textit{Herschel} is an ESA space observatory with science instruments provided by European-led Principal Investigator consortia and with important participation from NASA.} Exploitation of Local Galaxy Andromeda (HELGA)}
\subtitle{VII. A \textit{SKIRT} radiative transfer model and insights on dust heating}

\author{S. Viaene\inst{\ref{inst-UGent}}
\and M. Baes \inst{\ref{inst-UGent}} 
\and A. Tamm \inst{\ref{inst-Tartu}} 
\and E. Tempel \inst{\ref{inst-Tartu}} 
\and G. Bendo\inst{\ref{inst-Manch}} 
\and J. A. D. L. Blommaert\inst{\ref{inst-KUL}, \ref{inst-VUB}} 
\and M. Boquien\inst{\ref{inst-Chili}} 
\and A. Boselli\inst{\ref{inst-Marseille}} 
\and P. Camps\inst{\ref{inst-UGent}} 
\and A. Cooray\inst{\ref{inst-Irvine}} 
\and I. De Looze\inst{\ref{inst-UGent},\ref{inst-UCL}} 
\and P. De Vis\inst{\ref{inst-UGent}} 
\and J. A. Fernández-Ontiveros\inst{\ref{inst-Rome}} 
\and J. Fritz\inst{\ref{inst-UNAM}} 
\and M. Galametz\inst{\ref{inst-Garching}} 
\and G. Gentile\inst{\ref{inst-UGent}} 
\and S. Madden\inst{\ref{inst-CEA}} 
\and M. W. L. Smith\inst{\ref{inst-Cardiff}} 
\and L. Spinoglio\inst{\ref{inst-Rome}} 
\and S. Verstocken\inst{\ref{inst-UGent}} 
}

\institute{Sterrenkundig Observatorium, Universiteit Gent, Krijgslaan 281, B-9000 Gent, Belgium \label{inst-UGent} \\
\email{sebastien.viaene@ugent.be}
\and
Tartu Observatory, Observatooriumi 1, 61602 T\~oravere, Estonia \label{inst-Tartu}
\and
UK ALMA Regional Centre Node, Jodrell Bank Centre for Astrophysics, School of Physics and Astronomy, University of Manchester, Oxford Road, Manchester M13 9PL, UK \label{inst-Manch}
\and
KU Leuven, Instituut voor Sterrenkunde, Celestijnenlaan 200D 2401, 3001 Leuven, Belgium \label{inst-KUL}
\and
Vrije Universiteit Brussel, Department of Physics and Astrophysics, Pleinlaan 2, 1050 Brussels, Belgium \label{inst-VUB}
\and
Unidad de Astronomía, Fac. Cs. Básicas, Universidad de Antofagasta, Avda. U. de Antofagasta 02800, Antofagasta, Chile \label{inst-Chili}
\and
Laboratoire d'Astrophysique de Marseille – LAM, Université d'Aix-Marseille \& CNRS, UMR7326, 38 rue F. Joliot-Curie, F-13388 Marseille Cedex 13, France \label{inst-Marseille}
\and
Department of Physics and Astronomy, University of California, Irvine CA 92697 USA \label{inst-Irvine}
\and
Department of Physics and Astronomy, University College London, Gower Street, London WC1E 6BT, UK\label{inst-UCL}
\and
Istituto di Astrofisica e Planetologia Spaziali (INAF-IAPS), Via Fosso del Cavaliere 100, I-00133 Roma, Italy \label{inst-Rome}
\and
Instituto de Radioastronom\'\i a y Astrof\'\i sica, IRAf, UNAM, Campus Morelia, A.P. 3-72, C.P. 58089, Mexico \label{inst-UNAM}
\and
European Southern Observatory, Karl-Schwarzschild-Str. 2, D-85748 Garching-bei-München, Germany \label{inst-Garching}
\and
Laboratoire AIM, CEA, Université Paris Diderot, IRFU/Service d'Astrophysique, Bat. 709, F-91191 Gif-sur-Yvette, France \label{inst-CEA}
\and
School of Physics and Astronomy, Cardiff University, Queens Buildings, The Parade, Cardiff CF24 3AA, UK \label{inst-Cardiff}
}

\abstract{The radiation of stars heats dust grains in the diffuse interstellar medium and in star-forming regions in galaxies. Modelling this interaction provides information on dust in galaxies, a vital ingredient for their evolution. It is not straightforward to identify the stellar populations heating the dust, and to link attenuation to emission on a sub-galactic scale. Radiative transfer models are able to simulate this dust-starlight interaction in a realistic, three-dimensional setting. We investigate the dust heating mechanisms on a local and global galactic scale, using the Andromeda galaxy (M31) as our laboratory.

We perform a series of panchromatic radiative transfer simulations of Andromeda with our code SKIRT. The high inclination angle of M31 complicates the 3D modelling and causes projection effects. However, the observed morphology and flux density are reproduced fairly well from UV to sub-millimeter wavelengths. Our model reveals a realistic attenuation curve, compatible with previous, observational estimates. We find that the dust in M31 is mainly ($91 \%$ of the absorbed luminosity) heated by the evolved stellar populations. The bright bulge produces a strong radiation field and induces non-local heating up to the main star-forming ring at 10 kpc.

The relative contribution of unevolved stellar populations to the dust heating varies strongly with wavelength and with galactocentric distance. The dust heating fraction of  unevolved stellar populations correlates strongly with $NUV-r$ colour and specific star formation rate. These two related parameters are promising probes for the dust heating sources at a local scale.}


\titlerunning{HELGA VII: dust heating in M31}
\authorrunning{S. Viaene}

\maketitle

\section{Introduction}

Starlight in galaxies is processed by dust grains through absorption and scattering. On average, about one third of starlight in normal spiral galaxies is attenuated by dust \citep{Popescu2002,Skibba2011,Viaene2016}. Dust resides around new stars in their birth clouds as well as in the diffuse interstellar medium (ISM). A Milky Way-like dust mixture is highly efficient in scattering and absorbing UV radiation, but also significant fractions of optical and sometimes even near-IR (NIR) light. All the energy that is absorbed by the dust is reprocessed and emitted at longer wavelengths.

In order to understand the processes governing the observed dust-starlight interplay, observations at all wavelengths are necessary. This allows the investigation of how the UV and optical energy from stars is converted into far-IR (FIR) and sub-millimeter (submm) emission. Such knowledge is important since FIR/submm emission is a useful tool to study galaxies both near and far; the amount of dust obscuration roughly follows the star formation rate density in cosmic time \citep[see e.g.][]{Daddi2005, Gruppioni2013, Madau2014}.  A correct understanding of dust heating allows the separation of dust into multiple temperature components. This in turn results in better dust mass estimates. Accurate dust masses can then be used to estimate the total ISM content in distant objects for which H\textsc{i} content cannot be observed \citep[see e.g.][]{Corbelli2012, Eales2012, Scoville2014, Groves2015}. 

Nearby galaxies are good test laboratories for such dust heating studies. Their dust heating mechanisms have been studied through various techniques such as correlations with star formation rate (SFR) indicators and stellar mass tracers \citep{Galametz2010, Boquien2011, Bendo2012, Foyle2013, Hughes2014, Bendo2015} or through panchromatic energy balance SED fitting \citep{Groves2012, Aniano2012, Dale2012, MentuchCooper2012, Ciesla2014, RemyRuyer2015, Boquien2016}. While the situation differs from galaxy to galaxy, the current understanding is that the cold dust in most galaxies is heated by the evolved stellar populations which feed the general interstellar radiation field (ISRF). The warmer dust is then predominantly heated by the young and new stars, through strong UV radiation. In certain cases, even active galactic nuclei (AGN) can contribute to the dust heating \citep[see e.g.][]{Wu2007, Kirkpatrick2012, Schneider2015}. 

Usually, these investigations provide a qualitative view on the dominant heating source. In starbursts and galaxies with active star formation, dust is predominantly heated by young stellar populations. For early-type galaxies, little research has been conducted to estimate the dust heating sources. Based on their low star formation rates, one would expect that the evolved populations are dominant in these systems \citep[see e.g.][]{Smith2012b, Boquien2014}. But other heating sources are possible, for example X-rays from hot halo gas \citep{Natale2010}. Most galaxies, however, fall in the ambiguous intermediate regime, where both evolved and young stellar populations contribute to the dust heating \citep{Bendo2015}. It is dangerous to infer properties like dust mass and SFR from FIR/submm data alone in these systems. 

A powerful method to investigate dust heating processes is through dust continuum radiative transfer (RT) simulations, which mimic the transport of radiation through a dusty medium. For a review of this method, we refer the reader to \citet{Steinacker2013}. Such simulations treat the dust-starlight interaction in 3D, allowing realistic geometries, multiple viewing angles and non-local heating. But it also comes with a large computational cost and the need for ``simple'', well-behaved targets to model. Therefore, the most common application is in models of edge-on spiral galaxies \citep[see e.g.][]{Popescu2000, Misiriotis2001, Bianchi2008, Baes2010, Popescu2011, MacLachlan2011, DeLooze2012b,DeGeyter2015} or early-type galaxies \citep{DeLooze2012a, Viaene2015}. 

Through consistent treatment of primary emission, scattering, absorption and thermal re-emission, dust heating mechanisms can be investigated. A general result of these studies is that starlight absorption by diffuse dust usually fails to heat this dust to sufficient levels to match the observed FIR/submm emission. Often, obscured star formation is necessary to balance the energy between UV/optical (absorption) and FIR/submm (emission).

In a recent attempt to quantify the dust heating mechanisms, \citet{DeLooze2014} performed a detailed radiative transfer simulation of M51. Based on observed images, they closely mimic the distribution of both stars and dust, and then perform an accurate treatment of the dust-starlight interaction. They find that the contribution of evolved stars to the thermal re-emission is significant and varies with location and observed IR wavelength. In the mid-IR (MIR), the evolved stellar populations contribute about $10\%$ to the total dust heating from primary stellar emission, while young stars consequently contribute $90 \%$. This changes at submm wavelengths, where the contribution of evolved stars is $\sim40\%$. There is also a significant enhancement of the contribution by young stars in the spiral arms (>$80\%$) with respect to the inter-arm regions (<$60\%$).

In this paper, we apply this novel technique to the Andromeda galaxy (M31). It is the most massive object in our Local Group and close to our own Milky Way. In the optical, the galaxy is dominated by its bright bulge \citep{Tempel2010}. The smooth disk is intersected by several dark lanes where dust obscures the starlight. The dark dust patches coincide with bright emission in the FIR and submm \citep{Fritz2012, Groves2012}. On the other side of the spectrum, in the UV, a striking similarity with the FIR morphology is observed \citep{Thilker2005}. The morphological (anti-)correlations are a manifestation of intricate interactions between dust and starlight. It is also evident that the observed features point at a complex 3D structure. Looking at M31 as it is projected on the sky inevitably leads to parameters which are summed or averaged along the line of sight. A 3D model of the sources and sinks for radiation in M31 can help to understand our closest neighbour beyond the standard line-of-sight quantities. It grants us the opportunity to investigate the morphology from different viewing angles and its influence on the spectral energy distribution (SED), the attenuation law and the processes of scattering, absorption and re-emission by dust.

This endeavour sets new challenges to the radiative transfer modelling as M31 is much larger and has a higher inclination angle than M51 (in this work, we will use $i = 77.5 \degree$, which is the average value for the disk of M31 as derived from H\textsc{i} data by \citealt{Corbelli2010}). On the other hand, M31 is an obvious choice for studying dust heating mechanisms in great detail. Even for the hardest accessible wavelengths, the spatial resolution is still <140 pc along the major axis. Dust heating mechanisms in M31 have been investigated previously using UV-FIR energy balance SED fitting \citep{Montalto2009, Groves2012, Viaene2014}, using FIR/submm SED fitting by \citet{Smith2012} and \citet{Draine2014}, and through FIR colours \citep{PlanckM31}. All of these methods show that the emission from evolved stellar populations are the main contributor to the dust heating, especially at wavelengths beyond 160~$\upmu$m. However, quantifying the contributions of the different heating sources and their wavelength dependency remains difficult. Our study approaches this problem from a new perspective (radiative transfer simulations), and will provide quantitative measures related to dust heating in this galaxy.

In Sect.~\ref{sec:data} of this paper we describe the panchromatic dataset we use and core data products we derive from it. Our model and all its components are outlined in Sect.~\ref{sec:methods} and validated in Sect.~\ref{sec:validation}. We present a 3D view of Andromeda in Sect.~\ref{sec:3dview} and a dust heating analysis in Sect.~\ref{sec:heating}. We present our main conclusions in Sect.~\ref{sec:conclusions}.

\section{A panchromatic dataset of M31} \label{sec:data}

The data used in this work is the same as for our previous study \citep[][hereafter HELGA IV]{Viaene2014}. An elaborate description of the data can be found in that paper, we give a brief overview here.
The panchromatic dataset is a combination of dedicated observing campaigns using ground based and space telescopes. In the UV domain, M31 was observed by GALEX \citep{galex} in the FUV and NUV bands \citep{Thilker2005}. Optical data were acquired from SDSS \citep{sdss} and mosaicked by \citet{Tempel2012}, taking special care of the background variations. The result is a large (2.5$\degree$ x 8$\degree$) field of view in the $ugriz$ bands. In the near and mid-infrared, we rely on observations from \textit{Spitzer} \citep{Spitzer} using the IRAC instrument \citep{IRAC} and WISE \citep{WISE}. Andromeda was imaged in all four IRAC bands and is described in \citet{Barmby2006}. High-quality mosaics of the WISE observations were created in all four bands by the WISE Nearby Galaxy Atlas team \citep{Jarrett2013}. In the FIR regime, we use observations from the MIPS \citep{MIPS} instrument aboard \textit{Spitzer}. Due to resolution and sensitivity restrictions, we only use the maps at 24 and 70~$\upmu$m from \citet{Gordon2006}. 

Most recently, we completed our dataset with FIR/submm imaging using the \textit{Herschel} Space Observatory \citep{Herschel}. The \textit{Herschel} Exploitation of Local Galaxy Andromeda (HELGA) observed the galaxy with PACS \citep{pacs} at 100 and 160~$\upmu$m and with SPIRE \citep{spire} at 250, 350 and 500~$\upmu$m. The PACS data were reduced using standard recipe with HIPE v12 and SCANAMORPHOS v24 \citep{Roussel2013}. The SPIRE data using HIPE v12 and a custom method for the baseline subtraction (BriGAdE, Smith et al. in prep.). An elaborate account on the data acquisition and reduction can be found in \citet[][HELGA I]{Fritz2012} and \citet[][HELGA II]{Smith2012}. The \textit{Herschel} data for M31 is now publicly available online\footnote{\url{http://irsa.ipac.caltech.edu/data/Herschel/HELGA}} in its newly reduced form.

In \citetalias{Viaene2014}, we have performed a series of data manipulations in order to bring this diverse dataset to a consistent set of images. First, foreground stars were masked based on UV and NIR colours. The masked areas were replaced by the local background, which often was the diffuse emission from M31 itself. Second, all images were convolved to match the SPIRE 500~$\upmu$m beam, which is the largest in our set. Finally, the frames were rebinned and regridded to match this beam size. The end products are a series of images with a pixel size of 36 arcsec (or 140 pc along the major axis of Andromeda) and with a beam of that same size. Throughout the paper, we use a distance to M31 of 785 kpc \citep{McConnachie2005}.

We work within a limited field of view to ensure good coverage at all wavelengths. The studied region is limited by an ellipse with center at $\alpha =$ 00:43:06.28 and $\delta =$ +41:21:12.22, semi-major axis of $1.425 \degree$, semi-minor axis of $0.400 \degree$ and a position angle of $38.1 \degree$. The corresponding physical scale is about 19.5~kpc along the major axis. This corresponds to the $B$-band D$_{25}$ radius \citep{deVaucouleurs1991} and encloses the star-forming ring and the brightest outer dust features. A few faint substructures beyond $25$ kpc are not included in this field-of-view. This should not affect the dust heating analysis, which is limited to the inner $20$ kpc of the galaxy.

Although foreground stars and background galaxies are masked in the images by \citetalias{Viaene2014}, M32 and NGC 205 are not. Both dwarf galaxies are satellites of Andromeda and M32 in particular lies in front of our area of interest. This object is thought to have played an important role in the recent history of Andromeda due to a head-on collision \citep[see e.g.][]{Block2006,Gordon2006, Dierickx2014}. M32 is currently not part of the disk of Andromeda and its emission will influence our measurements of the main galaxy itself. We therefore mask out this dwarf companion where it is detected (from the FUV to the 24~$\upmu$m band). To do that, we fit a 2nd order 2D polynomial to the pixels surrounding M32 and replace the emission from M32 by the interpolated local background. 

Finally, we use integrated fluxes from IRAS \citep{IRAS} and Planck \citep{Planck} as additional constraints on the global SED of M31. IRAS fluxes at 12, 25, 60 and 100~$\upmu$m were taken from \citet{Miville2005}. We use the Planck 350, 550 and 850~$\upmu$m fluxes derived by \citet{PlanckM31}. Most of the images described above will serve as observational constraints on the model. However, we use a handful of images as an input for our model construction. They are assumed to trace different components of M31 and are discussed in Sect.~\ref{subsec:modsetup}.

\section{Constructing a 3D model} \label{sec:methods}

\subsection{Radiative transfer simulations}

We make use of SKIRT\footnote{SKIRT is publicly available at \url{http://skirt.ugent.be}} \citep{Baes2011,Camps2015}, an advanced dust continuum RT code that uses the Monte Carlo approach. The code allows panchromatic RT simulations using efficient dust grids \citep{Camps2013, Saftly2014}, a vast suite of possible geometries and geometry decorators \citep{Baes2015} and was recently updated with a user-friendly interface. It supports several dust models, and a full treatment of stochastically heated dust grains \citep{Camps2015b}.

A key feature for this investigation is the input of a 2D FITS image as a possible geometry, first demonstrated by \citet{DeLooze2014}. It uses the relative surface brightness in the image pixels to set up a 2D density distribution in the virtual 3D space. This density distribution can then be deprojected using a simple $\cos(i)$ factor, with $i$ the inclination angle. To give the deprojected 2D geometry a 3D nature, the density is smeared out in the vertical direction according to an exponential profile with a vertical scale height. This powerful technique allows more realistic stellar geometries such as asymmetric features or clumpy and filamentary structures. For M31 in particular, with its highly disturbed and clumpy rings \citepalias{Fritz2012}, this is a big step forward. Of course, the deprojection of a highly inclined galaxy such as M31 ($i = 77.5 \degree$) is always degenerate. Brighter spots in the disk will be smeared out in the direction of deprojection and the assumption of an exponential vertical profile will again smear the bright spots in the vertical direction. This technique does, by construction, conserve flux during the conversion from 2D to 3D. 

\subsection{Model set-up} \label{subsec:modsetup}

The radiative transfer model of Andromeda contains several stellar components and two dust components. The details of our model setup are outlined in Appendix~\ref{subsec:modcomponents}. We give a brief overview here. The components are summarized in Table~\ref{tab:components}.

\begin{table*} 
\caption{Overview of the different stellar populations (SP) and dust component in our model. We used an inclination angle of $77.5\degree$ to deproject the input images.}
\label{tab:components}
\centering     
\begin{tabular}{>{\raggedright\arraybackslash}m{3.4cm}>{\raggedright\arraybackslash}m{4.2cm}>{\raggedright\arraybackslash}m{2.7cm}>{\raggedright\arraybackslash}m{4.0cm}>{\raggedright\arraybackslash}m{2.2cm}}
\hline
\hline
Component & 2D geometry & vertical dimension & SED template & Normalization\\
\hline
\multicolumn{5}{l}{\textit{Bulge}} \\
Evolved SP (12 Gyr)	& \multicolumn{2}{c}{3D Einasto profile}	& \citet{Bruzual2003} & 3.6~$\upmu$m \\

\hline
\multicolumn{5}{l}{\textit{Disk}} \\
Evolved SP (8 Gyr)	& Corrected\tablefootmark{a} 3.6~$\upmu$m image & Exponential profile	& \citet{Bruzual2003} & 3.6~$\upmu$m \\
Young SP (100 Myr)	& Corrected\tablefootmark{b} $FUV$ image & Exponential profile	& \citet{Bruzual2003} & $FUV$ \\
Ionizing SP (10 Myr)	& Corrected\tablefootmark{b} 24~$\upmu$m image & Exponential profile	& MAPPINGS III\tablefootmark{c} & $FUV$ \\

\hline
\multicolumn{5}{l}{\textit{Dust}} \\
Interstellar dust & MAGPHYS dust mass map\tablefootmark{d} & Exponential profile	& THEMIS\tablefootmark{e} dust mix & Total dust mass \\

\hline
\end{tabular}
\tablefoot{
\tablefoottext{a}{The Einasto bulge was first subtracted from the image.}
\tablefoottext{b}{The contribution from the evolved stars was first subtracted from the image.}
\tablefoottext{c}{The SED templates for obscured star formation from \citet{Groves2008}.}
\tablefoottext{d}{The dust mass map derived from the pixel-by-pixel MAGPHYS energy balance SED models from \citet{Viaene2014}.}
\tablefoottext{e}{\citet{Jones2013}.}
}
\end{table*}

To keep the balance between a realistic model and computational feasibility, the number of components has to be limited. We split the stars in M31 in 4 components: an evolved bulge population (12 Gyr), an evolved disk population (8 Gyr), a young disk population (100 Myr), and an population of ionizing stars (10 Myr). The dust in M31 is split in an interstellar component, and a small fraction of dust around the ionizing stellar component. All stellar components have two aspects: geometrical distribution in 3D space, and a panchromatic SED.

The geometrical distributions are based on observed images. For the evolved populations, we take an analytical 3D \citet{Einasto1965} profile for the bulge. This bulge is subtracted from the IRAC 3.6~$\upmu$m image to obtain the disk as seen on the sky. To make the disk 3-dimensional, it is deprojected, and a vertical dimension is added in the form of an exponential surface brightness profile. The young stellar component is based on the $FUV$ image of M31, with a flux correction for UV emission from evolved stars. The ionizing stellar component is based on the MIPS 24~$\upmu$m image, also corrected for flux originating from evolved stars. The latter two observed images are also deprojected and a vertical exponential profile is added to obtain a 3D geometry. However, the vertical scale height of the young and ionizing component is set to be much smaller than for the evolved stellar disk. SEDs are assigned to the distributions, and normalized to the flux in a single waveband, see Table~\ref{tab:components}.

The dust around stars in their birth clouds is embedded as a subgrid recipe in the SED template for the ionizing stellar population. It thus follows the distribution of this stellar population. The distribution for the interstellar dust is derived from the dust mass map from \citet{Viaene2014}. This map is again deprojected and a vertical exponential profile is added to create a 3D distribution. The optical properties of the dust are set according to the THEMIS dust mix model \citep{Jones2013}.

The dust grid in our simulations is constructed using a hierarchical binary tree. The cells are built by splitting the parent cell in turns in the x, y and z direction. Cells are not further split when their dust mass fraction is below $10^{-6}$. For more information about this dust grid, we refer the reader to \citet{Saftly2014}. The grid is automatically constructed and has 1.45 million dust cells for the full M31 model. Note that these cells are not equal in size. Their volume ranges between (30-5000 pc)$^3$, with a median value of (89 pc)$^3$.

\subsection{Model optimization} \label{subsec:optimization}

Table~\ref{tab:parameters} gives an overview of the main parameters that have to be set in the simulation. For a detailed account on their physical meaning and the initial guess values, we refer the reader to Appendix~\ref{subsec:modcomponents}.

In order to find a set of parameters that provides a good working model, we fix all but three parameters: the intrinsic FUV luminosities of the young and ionizing stellar components, and the total dust mass. For computational reasons, we can only explore a small phase space. We therefore make our initial guess values as accurate as possible (see Appendix~\ref{subsec:modcomponents}). For the dust mass, we search the parameter space within a $\pm 25 \%$ range around our initial guess. The intrinsic FUV radiation of the young component ranges from $0.75-2.0$ times the initial guess value. The intrinsic FUV luminosities for the ionizing component ranges from 0.4 to $2.6 \times 10^8 \; L_\odot$.

\begin{table*} 
\caption{Overview of the parameters in our model. We used an inclination angle of $77.5\degree$ to deproject the input images. Most parameters remain fixed during the SED modelling. For our 3 variable parameters we give the range of our parameter grid. A detailed description of all parameters and the initial guess values can be found in Appendix~\ref{subsec:modcomponents}. Luminosities are in units of bolometric solar luminosity $L_\odot$.}
\label{tab:parameters}
\centering     
\begin{tabular}{>{\raggedright\arraybackslash}m{5.5cm}>{\raggedright\arraybackslash}m{2.5cm}>{\raggedright\arraybackslash}m{2.7cm}>{\raggedright\arraybackslash}m{2.9cm}>{\raggedright\arraybackslash}m{2.5cm}}
\hline
\hline
Description & Parameter & Initial Guess & Range & Best fit\\
\hline
\multicolumn{5}{l}{\textit{Evolved stars bulge (> 100 Myr)}} \\
Total luminosity at 3.6~$\upmu$m			& $\lambda L_\lambda^{\mathrm{bulge}}\, [L_\odot]$	& $1.53 \times 10^9$ 	& fixed &   \\
Central radius	of the Einasto profile	& $a_\mathrm{c}$ [kpc]	& 1.155					& fixed &   \\
Einasto structure parameter				& $N$					& 2.7					& fixed &   \\
Flattening factor						& $q$    				& 0.72					& fixed &   \\

\hline
\multicolumn{5}{l}{\textit{Evolved stars disk (> 100 Myr)}} \\
Total luminosity at 3.6~$\upmu$m			& $\lambda L_\lambda^{\mathrm{evolved}} \, [L_\odot]$	& $2.86 \times 10^9$ 		& fixed		&   \\
2D geometry of the evolved stars			& 2D image				& IRAC 3.6~$\upmu$m\tablefootmark{a}	& fixed & \\
Vertical scale height					& $h_\mathrm{z}$ [pc]	& 538								& fixed & \\

\hline
\multicolumn{5}{l}{\textit{Young stars (100 Myr)}} \\
Total intrinsic FUV luminosity 			& $\lambda L_\lambda^{\mathrm{young}} \, [L_\odot]$	& $0.93 \times 10^9$			& $(0.7-1.9) \times 10^9$ & $0.96 \times 10^9$ \\
2D geometry of the young stars			& 2D image				& GALEX FUV\tablefootmark{b}	& fixed & \\
Vertical scale height					& $h_\mathrm{z}$  [pc]	& 190						& fixed & \\

\hline
\multicolumn{5}{l}{\textit{Ionizing stars (10 Myr)}} \\
Total intrinsic FUV luminosity			& $\lambda L_\lambda^{\mathrm{ion.}} \, [L_\odot]$	& $0.4 \times 10^8$			 						& $(0.4-2.6) \times 10^8$ & $2.0 \times 10^8$  \\
2D geometry of the ionizing stars		    & 2D image							& MIPS 24~$\upmu$m\tablefootmark{c}	& fixed & \\
Vertical scale height					& $h_\mathrm{z}$ [pc]				& 190								& fixed & \\

\hline
\multicolumn{5}{l}{\textit{Interstellar Dust}} \\
Total dust mass							& $M_\mathrm{dust} \, [M_\odot]$	& $4.3 \times 10^7$ 	 	& $(3.2-5.4)\times 10^7$ & $4.7 \times 10^7$\\
2D geometry for the interstellar dust	& 2D image				& MAGPHYS map\tablefootmark{d}	& fixed & \\
Vertical scale height					& $h_\mathrm{z}$  [pc]	& 238							& fixed & \\
\hline
\end{tabular}
\tablefoot{
\tablefoottext{a}{The Einasto bulge was first subtracted from the image, see Sect.~\ref{subsec:OldGeometry}.}
\tablefoottext{b}{The contribution from the evolved stars was first subtracted from the image, see Sect.~\ref{sec:YoungComp}.}
\tablefoottext{c}{The contribution from the evolved stars was first subtracted from the image, see Sect.~\ref{sec:NewComp}.}
\tablefoottext{d}{The dust mass map derived from the pixel-by-pixel MAGPHYS energy balance SED models from \citet{Viaene2014}.}
}
\end{table*}

Using this parameter grid, we run a panchromatic RT simulation for each of the 1050 possible combinations of the free parameters. We compute the stellar emission and the effects of dust at 128 wavelengths between $0.05$ and $1000$~$\upmu$m. The wavelength grid is logarithmically spaced over this range, but has a finer spacing in the MIR range (1-30~$\upmu$m) to capture the aromatic features. The best fitting model SED is determined by the lowest $\chi^2$ between observed and model fluxes. The model fluxes were obtained by multiplying the model SED with the filter response curves and integrating the flux. The MIR dominates the $\chi^2$ when using a uniform weighing because there are many data points in this regime. However, to study the dust heating in M31, it is more important to adequately fit the UV attenuation and the dust emission peak. To ensure this, we lower the weights of the MIR data points to give priority to a good fit in the FUV and FIR/submm.

\begin{figure*}
	\centering
   	\includegraphics[scale=0.7]{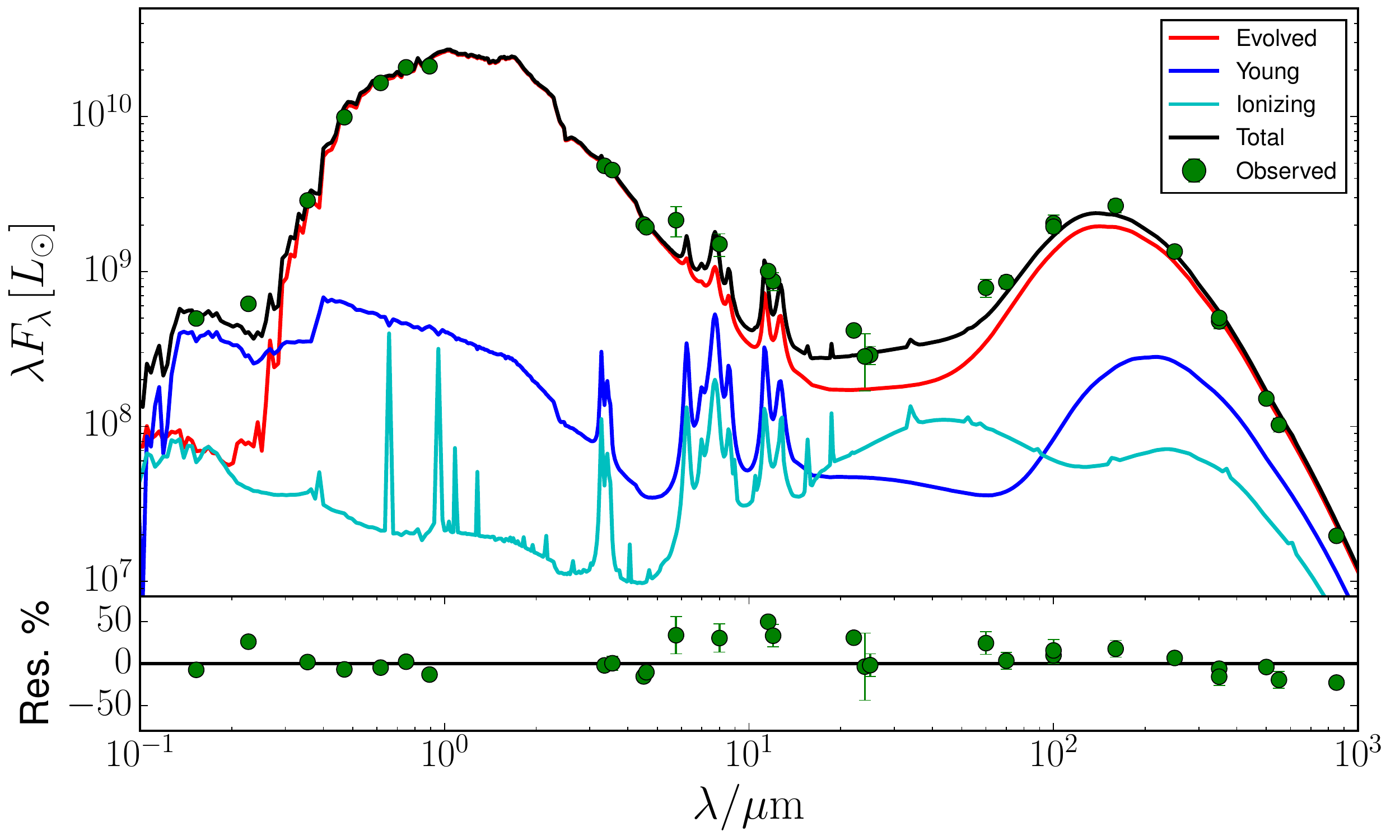}
	\caption{The panchromatic SED  of Andromeda. The black line is the best fitting RT model, run at high resolution (512 wavelengths). The green points are the observed integrated luminosities for M31. The red, blue and cyan lines represent the SEDs for simulations with only one stellar component: evolved, young and ionizing, respectively. The interstellar dust component is still present in these simulations. The bottom panel shows the residuals between the observations and the best fitting (black) model.}
	\label{fig:initialSED}
\end{figure*}

The best fitting model is re-simulated at higher resolution to extract images of higher signal-to-noise. The wavelength sampling was increased by a factor 4 to 512 wavelengths between 0.05 and 1000~$\upmu$m. Additionally, we use 20 times more photon packages ($20\times 10^6$ per wavelength in total) to accurately sample extinction, scattering, and emission. The SED of the high-resolution model is shown in Fig.~\ref{fig:initialSED}, together with the separate stellar components. To this end, we re-run the simulation each time using only one of the three stellar components (and still including the dust). Note that dust emission depends in a non-linear way on the absorbed energy, so the individual SEDs do not add up to the total SED. However, we can already learn some interesting trends on the energy balance of the galaxy.

The total SED of M31 is clearly dominated by the evolved stellar populations. They produce nearly all of the optical and NIR light, and are the strongest source in the MIR. They are also - indirectly - responsible for the bulk of the dust emission. We come back to this in Sect.~\ref{sec:heating}. Only in the UV domain, the contribution of the young (non-ionizing) stellar component is higher than that of the evolved stellar populations. The emission of the ionizing stellar populations is comparable to the evolved stellar populations in this domain. However, this is still about $5$ times lower than the young, non-ionizing component. 

The SED of the ionizing component has two emission peaks in the FIR/submm. This points to two very distinct dust components. The emission peak round $40 \, \upmu$m is associated with warmer dust, located inside the molecular clouds and heated by the ionizing stars. The second peak in the SED of the ionizing component occurs around $250 \, \upmu$m, indicating colder dust. This is emission from dust in the diffuse ISM, heated by radiation escaping the molecular clouds.

The emission from dust in the diffuse ISM peaks at different wavelengths in the SEDs of the different components. If only evolved stellar populations are included, the peak occurs at a much shorter wavelength ($\sim150 \, \upmu$m) than the peaks for the young ($\sim210 \, \upmu$m) and ionizing ($\sim250 \, \upmu$m) component. This indicates that the observed average dust temperature of Andromeda is mostly influenced by its evolved stellar populations. We perform a more quantitative analysis of the dust heating in Sect.~\ref{sec:ProjectedHeating}.

\section{Model validation} \label{sec:validation}

\subsection{Global SED} \label{subsec:globalSED}

Before investigating the 3D structure of M31 and the dust heating mechanisms in this galaxy, we perform a series of quality checks on the best fitting model. It has a dust mass of $M_\mathrm{dust} = 4.7 \times 10^7 M_\odot$. The intrinsic $FUV$ luminosity of the young and ionizing component is $\lambda L_\lambda^{\mathrm{young}} = 9.6 \times 10^8 L_\odot$ and $\lambda L_\lambda^{\mathrm{ion.}} = 2.0 \times 10^8 L_\odot$, respectively. This model SED is shown in the top panel of Fig.~\ref{fig:initialSED}, together with the observed fluxes. While the correspondence is generally good, there is still room for higher MIR/FIR luminosities to fit the observations in this area. However, that requires more FUV emission and/or more dust mass, which will affect the UV part of the SED. Since we tried these options in our modelling routine and did not get a lower $\chi^2$, we assume this situation is not preferable. 

The dust mass is on the high end of previous dust mass estimates (\citetalias{Fritz2012, Smith2012, Viaene2014}; \citealt{Draine2014}), but still consistent. One must note that a different dust model is assumed here. Despite comparable optical properties (tuned to fit the Milky Way extinction curve), the composition and grain density can differ significantly. This can lead to a difference of a factor 2-3 \citep{Jones2013,RemyRuyer2015,Dalcanton2015, Planck2016}. 

Our other two free parameters will have a significant influence on the SFR (both from UV or FIR measurements). We therefore `observe' our model SED and apply several recipes from \citet{Kennicutt2012} to convert flux to a SFR. The resulting values for the SFR are summarized in Table~\ref{tab:sfr}. For the $FUV$ and $NUV$ estimates, the intrinsic UV fluxes were used to avoid the problem of dust attenuation. The SFR values derived from  single band fluxes lie close to each other. They are well in line with previous estimates of the SFR in M31, which lie between 0.2 and 0.4 $M_\odot \mathrm{yr}^{-1}$ (see e.g. \citealt{Tabatabaei2010}; \citetalias{Viaene2014}; \citealt{Rahmani2016} or \citealt{Ford2013} (HELGA III)). 

A caveat to using only $24 \, \upmu$m or $70 \, \upmu$m flux is that it assumes all UV radiation from star formation is absorbed by dust. This is not the case for M31, so these bands naturally underestimate the SFR. We therefore also apply the hybrid SFR estimator from \citet{Leroy2008}, which combines the observed $FUV$ with the observed $24 \, \upmu$m flux. This recipe captures both the obscured and unobscured star formation. We find a value of 0.31 $M_\odot \mathrm{yr}^{-1}$, which is significantly higher than the single band FIR tracers. However, it is still of the same order of magnitude and consistent with literature values for the SFR in Andromeda.

Offset from these values is the SFR derived from the total infrared light (TIR). This calibration gives a SFR rate which is roughly three times higher than the other estimates. This is most likely because the dust heating in M31 is not dominated by the young stellar populations, as we will show in Sect.~\ref{sec:heating}. TIR is therefore not a reliable tracer of SFR in Andromeda. In fact, even for galaxies with more star formation than M31, TIR can lead to an overestimated SFR \citep{Boquien2014}.

\begin{table} 
\caption{Star formation rates for M31, derived from synthetic broad band fluxes from our model SED. For the $FUV$ and $NUV$ estimates, the intrinsic UV fluxes were used to avoid the problem of dust extinction. Conversions are based on the \citet{Kroupa2003} initial mass function.}
\label{tab:sfr}
\centering     
\begin{tabular}{lll}
\hline
\hline
Band & SFR $[M_\odot \mathrm{yr}^{-1}]$ & Recipe \\
\hline
$FUV$ & 0.24 & a,b \\
$NUV$ & 0.31 & a,b \\
24~$\upmu$m	 & 0.23 & c \\
70~$\upmu$m	 & 0.19 & d \\
TIR ($3-1000 \upmu$m)	 & 0.92 & a,b \\
$FUV+24 \,\upmu$m & 0.31 & e \\
\hline
\end{tabular}
\tablefoot{
a) \citet{Murphy2011}, b) \citet{Hao2011}, c) \citet{Rieke2009}, d) \citet{Calzetti2010}, e) \citet{Leroy2008}.
}
\end{table}

\begin{figure*}
	\centering
   	\includegraphics[scale=0.6]{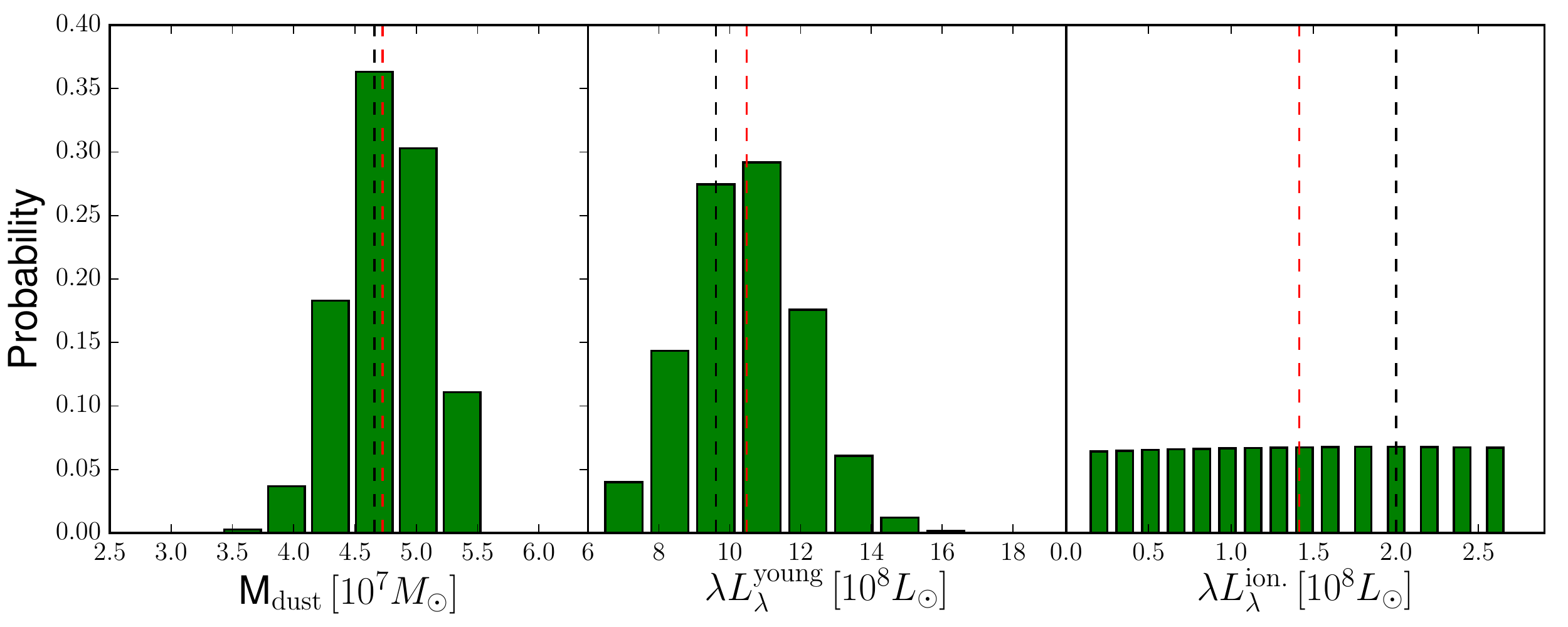}
	\caption{PDFs of the three free parameters in our model optimization: the total dust mass $M_\mathrm{dust}$ (left), the intrinsic $FUV$ luminosity from the young stellar component $\lambda L^\mathrm{young}_\lambda$ (middle) and the intrinsic $FUV$ luminosity from the ionizing stellar component $\lambda L^\mathrm{ion.}_\lambda$ (right). Black dashed lines are the parameter values for the best fitting model. Red dashed lines are the median (50th percentile) values from the PDFs.}
	\label{fig:plotPDFs}
\end{figure*}

To estimate the uncertainty on the free parameters, we construct probability distribution functions (PDFs). For each of the tested models, we compute the probability as proportional to $\exp(-\chi^2/2)$. For each parameter value we then add the probabilities and plot the PDFs in Fig.~\ref{fig:plotPDFs}. We find that dust mass and $\lambda L_\lambda^{\mathrm{young}}$ are fairly well constrained (the PDFs resemble a normal distribution). The $\lambda L_\lambda^{\mathrm{ion.}}$, on the other hand, has a more or less flat probability distribution and thus is poorly constrained. This is not so surprising since this parameter corresponds to constraining of the cyan curve in Fig.~\ref{fig:initialSED}. This normalization is fitted in a parameter range far below the luminosity of the old and young stellar components, making it difficult to find the optimal solution. The median (50th percentile) values that come from these PDFs are $M_\mathrm{dust} = 4.72_{-0.39}^{+0.36} \times 10^7 M_\odot$, $\lambda L_\lambda^{\mathrm{young}} = 10.5_{-1.6}^{+1.7} \times 10^8 L_\odot$ and $\lambda L_\lambda^{\mathrm{ion.}} = 1.4_{-0.6}^{+2.2} \times 10^8 L_\odot$. Here the upper and lower errors correspond to the 84th and 16th percentile of the PDFs, respectively.

\subsection{Attenuation law}

\begin{figure}
	\centering
   	\includegraphics[scale=0.35]{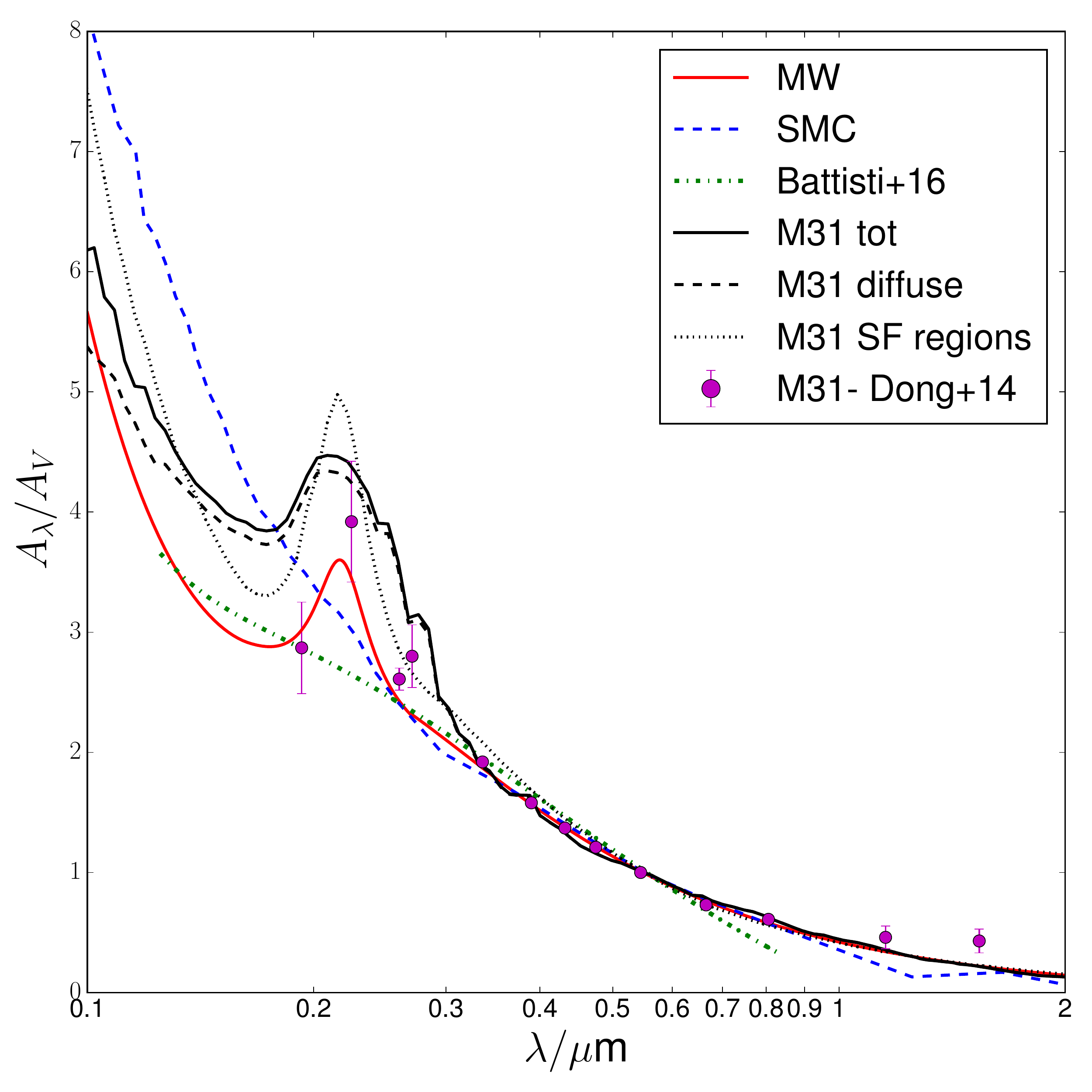}
	\caption{Attenuation laws from our model, normalized to the $V$-band attenuation. The black solid line is the global law. The dashed  line reflects the attenuation law for the diffuse dust and the dotted black line for the dust in star-forming regions. Several literature measurements are shown; red curve: MW curve for $R(V) = 3$ from \citet{Fitzpatrick2007}, blue dashed line: SMC bar region \citep{Gordon2003}, and green dash-dotted line: attenuation curve from \citet{Battisti2016}.}
	\label{fig:attCurve}
\end{figure}

We can accurately reconstruct the attenuation law in our model since we know the sources and sinks of the starlight. The attenuation curve for the best fit model is given in Fig.~\ref{fig:attCurve}, normalized to the $V$-band attenuation. We find a steadily increasing attenuation with decreasing wavelength, with on top of that a broad bump that peaks around $0.22\,\upmu$m. This curve is a combination of attenuation by diffuse dust as modelled by the dust mass map described in Sect.~\ref{subsec:DustGeometry}, and attenuation by dust in star-forming clouds. The latter is incorporated in the MAPPINGS III SED templates \citep{Groves2008}, which were used for the ionizing stellar component (see also Sect.~\ref{sec:NewComp}). The attenuation laws for both components are also shown in Fig.~\ref{fig:attCurve}. Note that the MAPPINGS attenuation law is directly provided by \citep{Groves2008}, while the diffuse dust attenuation curve is a combination of the THEMIS extinction law and the relative star-dust geometry.

The dust around new, ionizing stars contributes less than $0.1\%$ of the absorbed energy in the $V$ band. However, this ratio increases to $5 \%$ in the UV bands. 
The total attenuation curve is thus mainly shaped by the diffuse dust. The effect of the MAPPINGS III attenuation curve is visible in a slightly higher $NUV$ bump, and in a sharper increase at the shortest wavelengths. 

In Fig.~\ref{fig:attCurve}, we also compare against several literature attenuation curves: a) The average Milky Way extinction curve from \citet{Fitzpatrick2007}, converted to an attenuation curve using $R_V =3.001$ (see also their equation 10). b) The SMC bar region from \citet{Gordon2003}, again this is an extinction curve, converted to an idealized attenuation curve. c) The average attenuation law for $10000$ local ($z \lesssim 0.1$) star-forming galaxies from \citet{Battisti2016}. d) The average attenuation measurements from \citet{Dong2014} for the central 1' of M31. 

At NIR and optical wavelengths ($\lambda > 0.4\, \upmu$m) our attenuation curve follows the others almost perfectly. At shorter wavelengths, the curves start to diverge. The UV bump in our model curve is much broader than that of the MW. The broadness of the bump of this dust model is discussed in \citet{Jones2013}. It relates to the grain size distributions and their band gap properties. The discrepancy is amplified because the MW curve is an idealised attenuation curve, converted from an extinction law. This does not fully incorporate the large-scale spatial distribution of stars and dust. An additional complication in comparing attenuation curves is that their shape can depend on the resolution of the study \citep{Boquien2015}.

On the long wavelength side of the bump, our model is consistent with the empirical measurements of \citet{Dong2014}. Their UV bump does seem to be less pronounced. Unfortunately, they have no measurements shortward of 0.193~$\upmu$m. The broad UV bump can explain why our model has difficulties to reproduce the observed NUV flux (see Fig.~\ref{fig:initialSED}). 

Beyond the UV bump, our model curve steadily increases and falls somewhere between the MW and SMC idealised attenuation laws. In general, our model produces an attenuation law that is broadly in line with observations, indicating that our treatment of extinction and scattering yields realistic results.

\subsection{Images vs. observations} \label{sec:Residuals}

\begin{figure*}
	\centering
   	\includegraphics[width=1.0\textwidth]{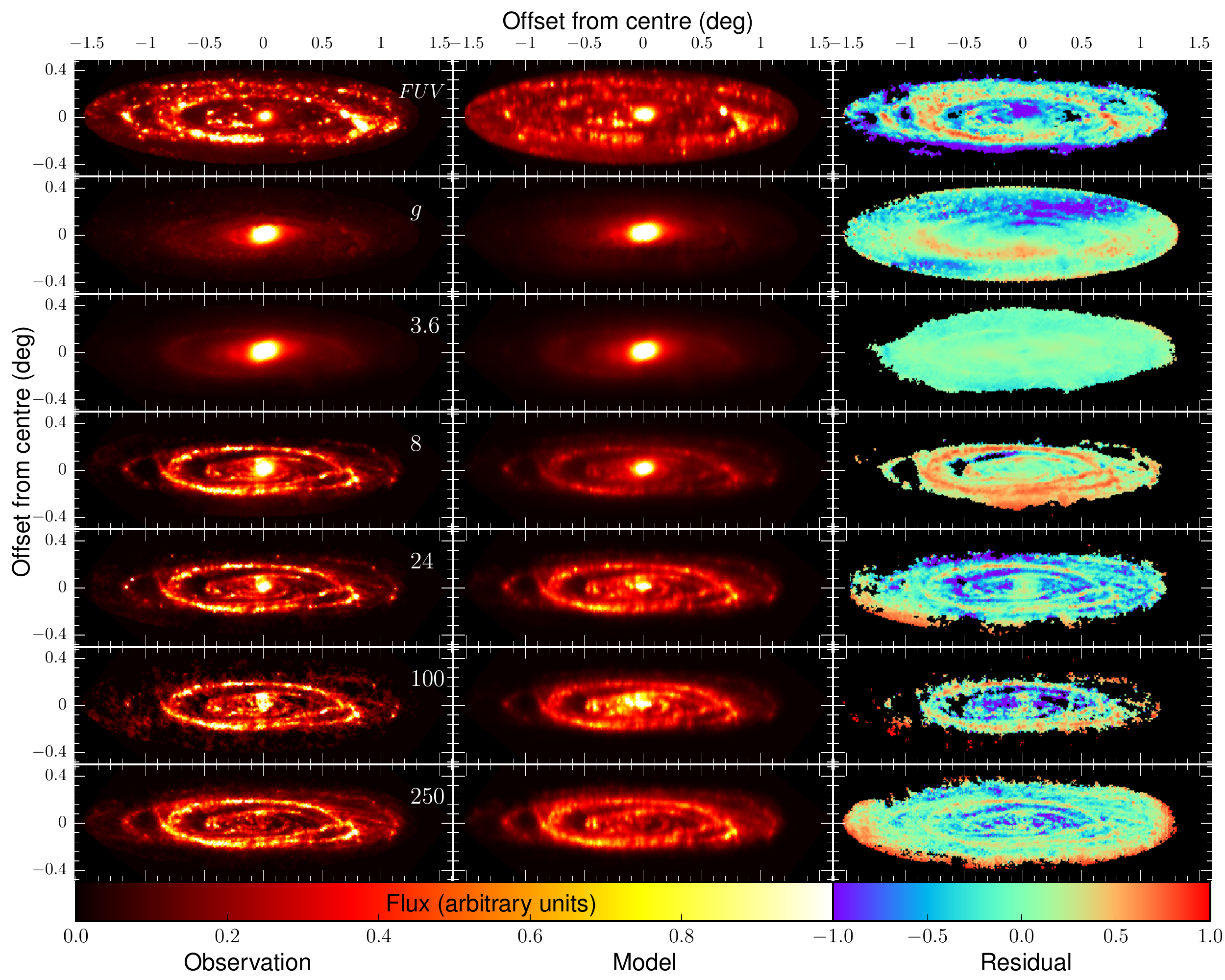}
	\caption{Comparison of the model with observations in selected wavebands. First column show the observed images, second column the model images. The 3rd column shows the residuals derived as in Eq.~\eqref{eq:residual}. From top to bottom row: $FUV$, $g$, 3.6~$\upmu$m, 8~$\upmu$m, 24~$\upmu$m, 100~$\upmu$m and 250~$\upmu$m.}
	\label{fig:residuals}
\end{figure*}

We now compare model and observations at a spatially resolved scale. Fig.~\ref{fig:residuals} shows a selection of important bands across the spectrum. We only show pixels with an observational SNR above 2. The easiest way to visualize the difference between observation (left column) and model (middle column) is through the residual image (right column):
\begin{equation} \label{eq:residual}
\mathrm{residual} = \frac{\mathrm{observation} - \mathrm{model}}{\mathrm{observation}}\text{.}
\end{equation}
A negative (positive) fraction means the model overestimates (underestimates) the observed flux.

The median (absolute value) deviation between model and observations across all bands is $22 \%$. Most of the model pixels fall within $50 \%$ of their observed counterpart. Some bands ($g$, 3.6~$\upmu$m and 250~$\upmu$m) have much smoother residuals, while others ($FUV$, 8~$\upmu$m, 24~$\upmu$m, 100~$\upmu$m) display clear structure.
The duality appears to be related to the appearance of M31 across wavelengths. In the NIR, Andromeda is a smooth galaxy without sharp features. The residuals in this regime are particularly regular. Bands in the MIR regime, with contributions of evolved stars, hot dust and aromatic features, show a very clumpy and ring-like Andromeda. The residuals in these bands exhibit strong features. 

In general, the model tends to underestimate the flux in the rings of M31 and overestimate the flux in the diffuse inter-ring regions by about $20\%$ on average. A first explanation for this can be the vertical dimension that was added to the model. Upon deprojecting the images, bright regions tend to be smeared out in the direction of deprojection. After that, the light is again smeared out in the vertical direction. In Appendix~\ref{sec:AltScaleHeights}, the effect of alternative scale heights for the disk components is investigated. However, we find that the effect is minor and cannot explain the bulk of the residual features.

A second aspect that could be responsible for at least a fraction of the discrepancy between observations and model is the limited resolution, and in particular the fact that we are not capable of modelling the star-forming regions in detail. Due to their intrinsic spherical shell geometry, these models emit isotropically and have a higher average attenuation per unit dust mass than models with a more asymmetric or clumpy distribution \citep{Witt1996, Witt2000, Varosi1999, Indebetouw2006, Whelan2011}. This effect is particularly relevant in the UV, where the ionizing stellar populations are relatively luminous and attenuation effects are pronounced.

As we discuss in the construction of our model (see Appendix~\ref{sec:YoungComp}), we probably underestimate the contribution of young stellar populations in the dustiest regions of the galaxy. Vice versa, we overestimate their contribution in the least attenuated regions. This could be another cause for the discrepancy observed in the residual maps in Fig.~\ref{fig:residuals}. In the model, the radiation field in the dustiest regions therefore does not hold enough $UV$ energy. As a result, the model underestimates the observed flux in these regions, and overestimates the flux in the more dust-free areas.

There may be a fourth effect contributing to the discrepancy. \citet{DeLooze2014} also observed residual features that could be associated with the spiral arms in their M51 model. They argue that there is a difference in both stellar populations and dust properties between arm and inter-arm regions. In \citetalias{Smith2012} it seemed that such variations caused differences in the dust emissivity index $\beta$ in their pixel-by-pixel modified black body fits of M31. However, a variable $\beta$ appears to correlate with $3.6 \, \upmu$m emission \citep{Kirkpatrick2014}. This suggests that variations in beta may not actually represent variations in dust emissivity but instead may represent a variation in dust heating sources. On the other hand, \citet{Galametz2012} found that allowing $\beta$ to vary can lead to unphysical dust temperatures in pixel-by-pixel energy balance SED fitting.

Nevertheless, it is not unlikely that spiral arms host different dust mixtures than inter-arm regions. While we allow for 3 stellar components in our model, we take the first order assumption of just one type of dust mixture for the entire galaxy. Grain size distributions may vary significantly depending on radiation field and ISM phase \citep[see e.g.,][]{Ysard2013,Ysard2015}. However, it would lead us too far to investigate the changing size distributions and grain properties in the different regions of M31. Furthermore, this would multiply the number of parameters in our model. We therefore choose to stay at the current level of complexity, while being aware of the caveats.

\subsection{FIR colours}

\begin{figure*}
	\centering
   	\includegraphics[width=0.8\textwidth]{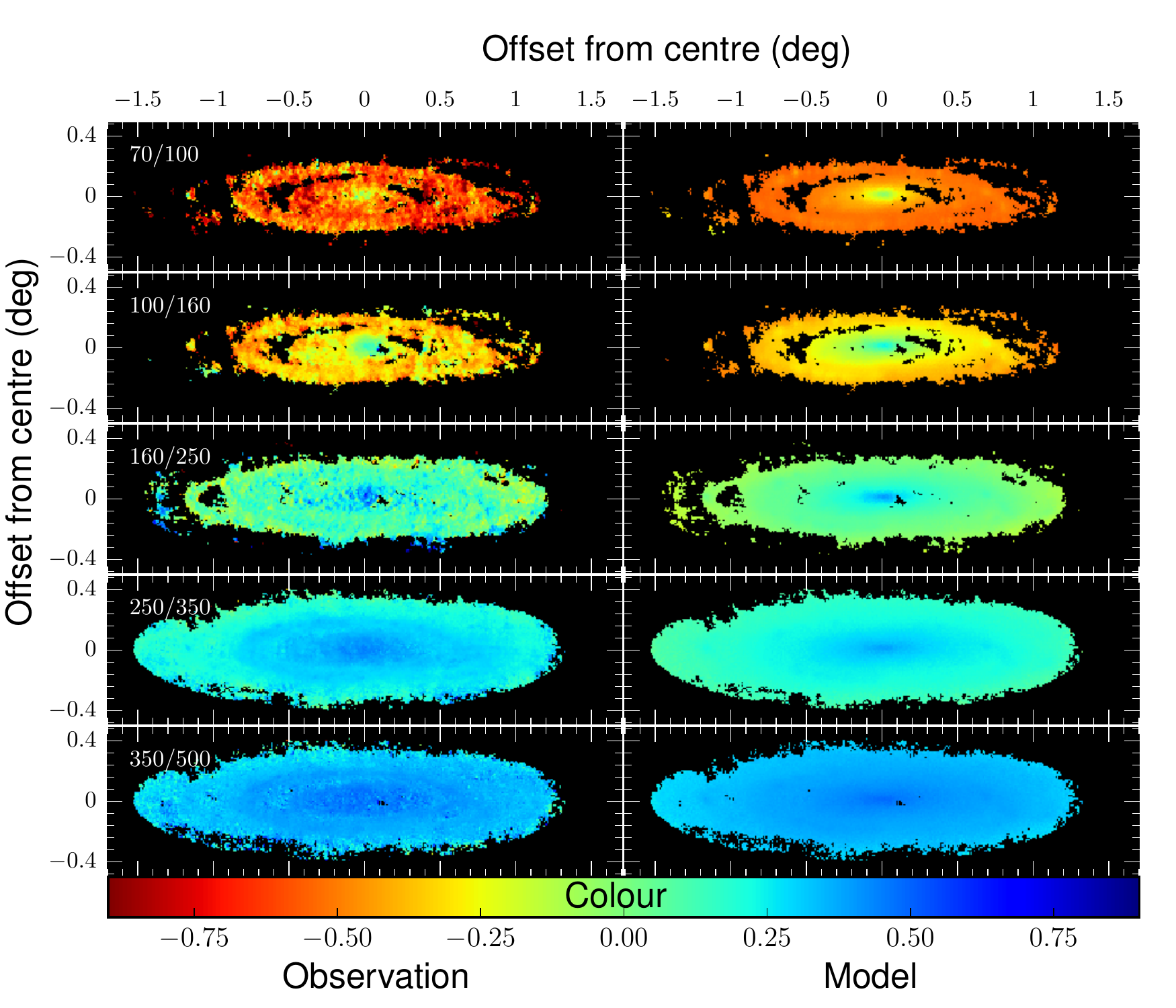}
	\caption{Observed (left) and model (right) colours tracing the FIR dust emission peak. A colour here corresponds to the flux ratio: $ \log(\mathrm{flux}_1/\mathrm{flux}_2)$, i.e. blue colours indicate a high flux ratio, red colours a low ratio.}
	\label{fig:FIRColours}
\end{figure*}

As a final check for our model, we compare subsequent FIR/submm colours with observed ones in Fig.~\ref{fig:FIRColours}. Note that we define colours as logarithmic flux ratios. We only considered pixels with  S/N > 3 as the PACS maps can have large uncertainties at low surface brightness. There is generally a good correspondence between model and observation. This was already evident from the global SED in Fig.~\ref{fig:initialSED}, where the cold dust peak is adequately reproduced. The colour maps, however, show that this is also the case on the resolved scale.

The observed $70/100$ and $100/160$ colours are more `clumpy' than in the model. This is mostly due to the low S/N in the MIPS and PACS observations. On average, however, the correspondence is satisfactory, although the observed maps are slightly redder. We chose not to use these colours for further analysis because of the higher uncertainties that come with it.

Of particular interest are the 160/250 and 250/350 surface brightness ratios, which cover the wavelength range where the transition of dust heating sources occurs. Consequently, they have been used in the past to identify the dominant dust heating sources \citep[see e.g.][and references therein]{Bendo2010, Boquien2011, Boselli2010, Boselli2012, Bendo2012, Bendo2015, PlanckM31}. There is a reasonable agreement in the morphology, with blue colours in the centre, and redder colours in the outskirts. The model maps are again smoother than the observed colour maps, but the difference is less pronounced than at shorter wavelengths. In Fig.~\ref{fig:initialSED}, we find that our model falls below the observation at 160~$\upmu$m, while it agrees with the 250~$\upmu$m flux. The model thus underestimates the 160/250 colour on the global scale. The resolved colour map indicates that this underestimation comes from the star-forming ring. The colours are bluer in the observed image. This is partially compensated by the centre and outskirts, where the model produces slightly redder colours than observed. The 250/350 colour is also underestimated and the offset is more general. The observed colour is bluer than the model in most of the pixels.

For completeness, we also present the $350/500$ colour. This is often neglected on a local scale because of the low angular resolution that comes with including the SPIRE $500\, \upmu$m band. Additionally, both the $350\, \upmu$m and $500\, \upmu$m band trace the Rayleigh-Jeans tail of the cold dust emission. It is therefore less informative than colours at shorter wavelengths. Again, observations and model are well in line with each other. The main deviations are the slightly redder outer regions in the observed colour map, but the effect is minor.

Taken together, the observed colours are fairly well reproduced by our model. This points at an adequate treatment of the dust physics, which is reassuring.

\section{A face-on view of M31} \label{sec:3dview}

\begin{figure*}
	\centering
   	\includegraphics[width=1.0\textwidth]{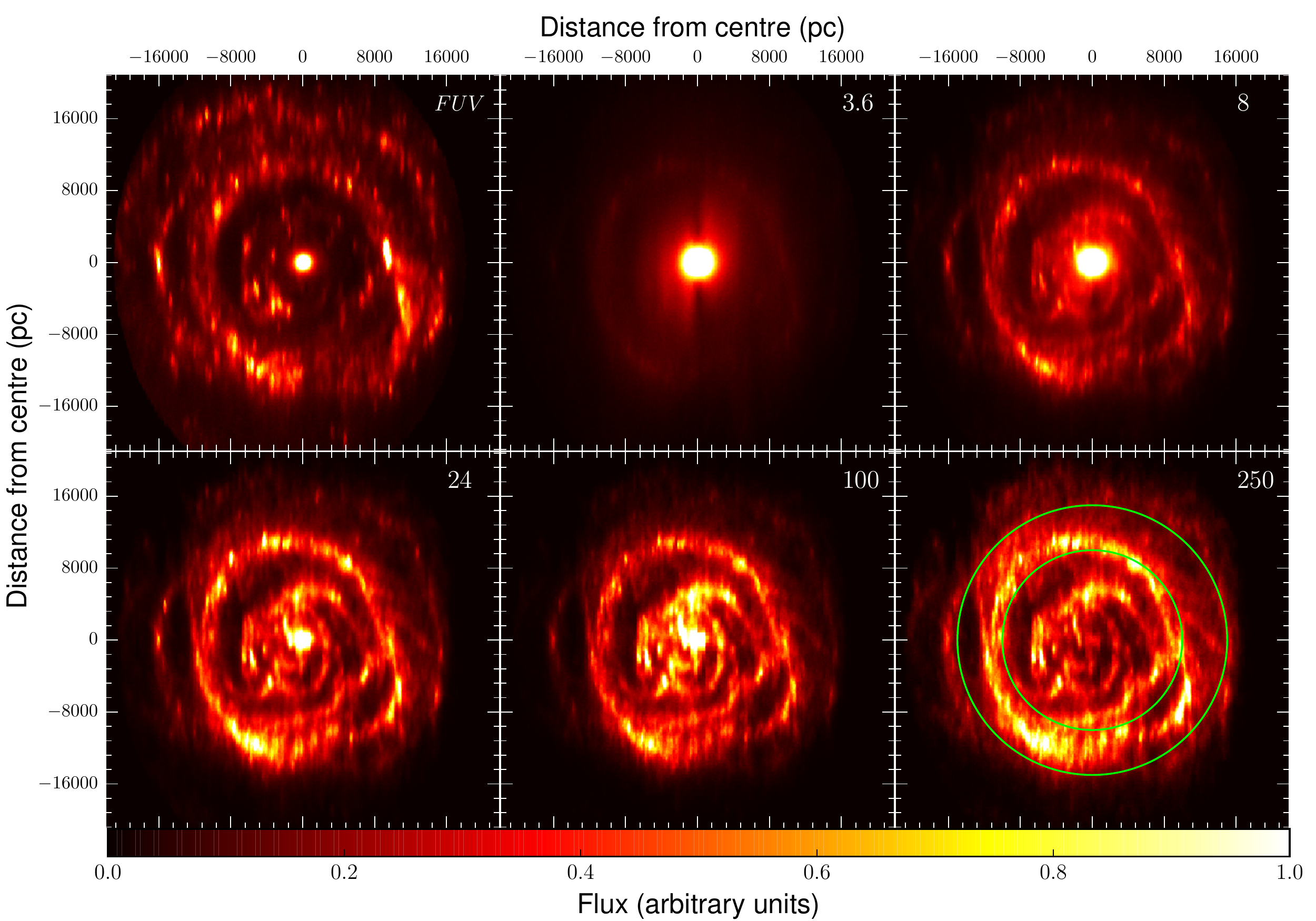}
	\caption{Face-on model images of M31 in the $FUV$, 3.6, 8.0, 24, 100 and 250~$\upmu$m. Colour scale is different for each image for clarity. The green circles on the 250~$\upmu$m image (bottom right) correspond to radii of 10 kpc and 15 kpc from the centre.}
	\label{fig:M31FaceOn}
\end{figure*}

The 3D RT simulations of Andromeda allow us to view the galaxy from different angles. Most informative is the face-on view shown in Fig.~\ref{fig:M31FaceOn} in six bands. The images highlight the diversity and complexity of the galaxy across the spectrum. In the $FUV$ band, the clumpy star formation regions are  structured in several rings and arcs. The bulge is more compact than in the optical, but still the brightest region. Most of the bright spots are elongated in the direction of the deprojection. This effect is inherent to the method and can't be corrected for. 

At 3.6~$\upmu$m the galaxy has a smooth disk with only the ring at 10 kpc as the reminder of the complex ring-arc morphology visible at 0.153~$\upmu$m. The bulge is large and dominates the NIR emission. A small gap can be observed in the middle of the image, along the vertical direction. This feature is an artefact of the bulge subtraction from the disk. Because we approximate the observed boxy bulge with an Einasto profile, some residual sidelobes remain in the disk. These sidelobes are smeared out in the direction of deprojection. This creates the two brighter features on the upper right and lower left of the vertical `gap' in the model 3.6~$\upmu$m image. 

The bulge is still the brightest feature  at 8~$\upmu$m.  Through the FIR and submm (24, 100  and 250~$\upmu$m images) the look of Andromeda gradually evolves.  The bulge becomes less dominant and the inner and outer rings turn brighter. The rings also exhibit a clear broadening going toward the submm as the extent of the diffuse dust becomes apparent at these wavelengths. While the bulge is still visible at 100~$\upmu$m, there is no sign of it at 250~$\upmu$m and M31 becomes one swirl of clumpy rings. 

\section{Dust heating mechanisms} \label{sec:heating}

We now aim to identify the dust heating sources and quantify their respective contribution. We will first look at the intrinsic 3D radiation field and link that to the heating of dust (Sect.~\ref{sec:3Dheating}). Then, we define and compute the dust heating parameters for the projected model on the sky (Sect.~\ref{sec:ProjectedHeating}). This way, we can link them to observational properties (Sect.~\ref{sec:heatingTracers}).

\subsection{3D analysis} \label{sec:3Dheating}

We can consider each dust cell in the model as a voxel (i.e. a 3D pixel) that processes stellar radiation. During the simulation, the absorbed luminosity $L_\lambda^\mathrm{abs}$ is stored in every cell and at every wavelength. Integrating over wavelength gives the total absorbed luminosity $L^\mathrm{abs}$ in each cell.

Such an approach does require a good sampling of the radiation field in each dust cell during the simulation. If only a few photon packages pass through a dust cell, the Monte Carlo noise will be high. This issue is not relevant when looking at the line-of-sight projected pixels, since many photon packages travel through the line-of-sight and provide sufficient sampling. To investigate the absorption for the dust cells, we run a separate set of simulations with more photon packages ($10^8$ per wavelength). The number of wavelengths is limited to 25, sampled logarithmically between 0.1 and 10~$\upmu$m, to compensate the computational cost that comes with shooting more photon packages. This wavelength range is adequate since the absorbed radiation mainly originates from stellar sources. We integrate the absorbed luminosity along the wavelength axis to compute $L^\mathrm{abs}$. The coarser sampling of the wavelength grid is compensated by the gain in signal-to-noise due to the integration. This method is similar to the formalism outlined in \citet{Natale2015}, equations (3)-(5). They start from the stellar radiation field to compute the total absorbed luminosity, whereas we start from the absorbed energy per cell. In a well-sampled simulation the two formalisms are equivalent.

We perform this simulation 4 times: including all stellar components yields $L^\mathrm{abs}_\mathrm{tot}$. Including only evolved, young or ionizing stellar populations yields $L^\mathrm{abs}_\mathrm{evolved}$, $L^\mathrm{abs}_\mathrm{young}$ and $L^\mathrm{abs}_\mathrm{ion.}$, respectively. $L^\mathrm{abs}_\mathrm{ion.}$ is the luminosity that originates from the MAPPINGS template SED, and is absorbed by the diffuse dust component. There is also an internally absorbed luminosity in the subgrid implementation of the MAPPINGS model for each dust cell. This can be seen as the absorbed energy from ionizing stars by dust in star-forming clouds $L^\mathrm{abs}_\mathrm{ion, SF}$ (as opposed to energy absorption due to diffuse ISM dust, $L^\mathrm{abs}_\mathrm{ion.}$). 

\begin{figure*}
	\centering
   	\includegraphics[scale=0.48]{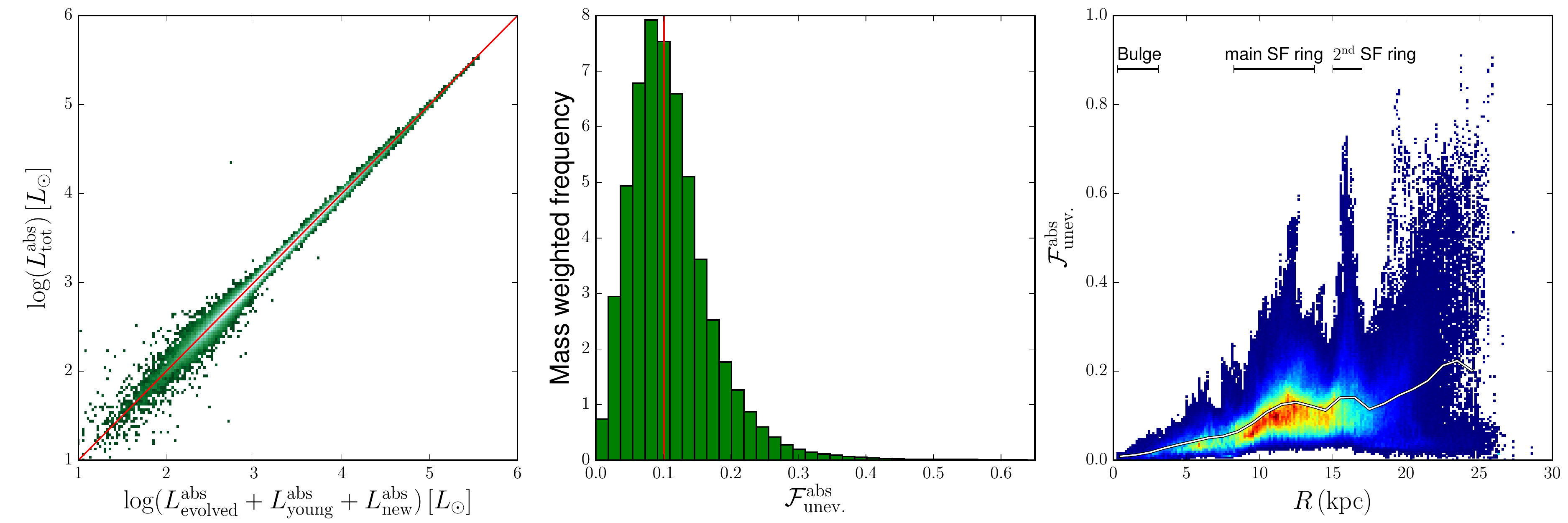}
	\caption{Left: 2D histogram comparing the integrated absorbed luminosity per dust cell including all stellar components $L_\mathrm{tot}^\mathrm{abs}$ with the same quantity, but as the sum of separate simulations including only evolved, young and ionizing stellar populations. Middle: Distribution of the bolometric absorption fraction of unevolved stellar emission per dust cell: $\mathcal{F}_\mathrm{unev.}^\mathrm{abs}$. The histogram is weighted by mass fraction of each dust cell and normalized. The red line indicates the median $\mathcal{F}_\mathrm{unev.}^\mathrm{abs} = 0.10$. Right: 2D histogram showing the radial distribution of $\mathcal{F}_\mathrm{unev.}^\mathrm{abs}$. The bins are weighted by the mass fraction of the dust cells. Red indicates a high number of data points, blue a low number. The spread at a fixed radius is a combination of dust cells at the same radius and with different vertical location in the disk. The white line is the weighted mean heating fraction per radius.}
	\label{fig:isrf}
\end{figure*}

To test the sampling of the radiation field in our simulations, we plot the sum of $L^\mathrm{abs}_\mathrm{evolved}$, $L^\mathrm{abs}_\mathrm{young}$ and $L^\mathrm{abs}_\mathrm{ion.}$ against the total absorption by the diffuse dust in Fig.~\ref{fig:isrf}. It is not necessary to count the internal absorption in the star-forming clouds here because the energy balance is internally secured in the subgrid model. The 1:1 line is plotted in red and shows a good correspondence between the three quantities. Only at low absorption luminosities, the variance starts to increase. This indicates that the radiation field in each dust cell is well sampled and we can separate the contribution from evolved, young and ionizing stellar populations to the total absorbed luminosity. We then define the contribution of young and ionizing stellar populations to the absorbed energy per dust cell as
\begin{equation} \label{eq:Labs}
\mathcal{F}_\mathrm{unev.}^\mathrm{abs} \equiv \frac{L^\mathrm{abs}_\mathrm{young}+L^\mathrm{abs}_\mathrm{ion.}+L^\mathrm{abs}_\mathrm{ion, SF}}{L^\mathrm{abs}_\mathrm{evolved}+L^\mathrm{abs}_\mathrm{young}+L^\mathrm{abs}_\mathrm{ion.}+L^\mathrm{abs}_\mathrm{ion, SF}},
\end{equation}
where we consider the contributions of the young and ionizing stellar populations together under the umbrella of `unevolved stellar populations'. Consequently, the absorbed luminosity fraction from the evolved stellar populations is  $\mathcal{F}_\mathrm{ev}^\mathrm{abs} = 1 - \mathcal{F}_\mathrm{unev.}^\mathrm{abs}$.

First, we take a global measure of the dust heating by summing, for each stellar component, the absorbed luminosities over the dust cells. Equation~\ref{eq:Labs} can then be applied to these global absorption luminosities to yield the global absorption fraction for the galaxy. We find that $\mathcal{F}_\mathrm{unev.}^\mathrm{abs} = 9 \%$ on a global scale for M31. In other words, $91 \%$ of the energy absorbed by dust originates from the evolved stellar populations. This is  a particularly high number (\citealt{DeLooze2014} found $37 \%$ for M51 using a comparable technique) and most likely dominated by strong radiation field of the bulge.

Second, we take a closer look at the dust heating of the individual dust cells in the middle and right panel of Fig.~\ref{fig:isrf}. The middle panel shows the distribution of $\mathcal{F}_\mathrm{unev.}^\mathrm{abs}$ for the dust cells. We weighted each dust cell with its mass fraction to account for the difference in size and dust content between the individual dust cells. The distribution peaks near the median value of $0.10$ and is skewed towards higher fractions. However, the contribution of unevolved stellar populations to the absorbed energy hardly exceeds $50\%$. This indicates that, even at the level of individual dust cells, the heating by evolved stellar populations dominates. 

The radial variation of the dust heating fraction can be investigated by plotting $\mathcal{F}_\mathrm{unev.}^\mathrm{abs}$ as a function of galactocentric radius for each dust cell. In the right panel of Fig.~\ref{fig:isrf}, this is shown for each dust cell, irrespective of its vertical coordinate. The 2D histogram is weighted by the mass fraction, and the weighted mean $\mathcal{F}_\mathrm{unev.}^\mathrm{abs}$ for each 1 kpc bin is shown as a white line. The contribution of the unevolved stellar populations is only a few percent in the inner few kpc, where the bulge is dominant. There is a gradual increase when moving to larger galactocentric radii. However, there is a large variation in $\mathcal{F}_\mathrm{unev.}^\mathrm{abs}$ for a fixed radius, especially at larger radii. Each radius contains dust cells from all around the galaxy, and from different vertical coordinates. The rings in M31 are also not concentric. The variance per radius therefore points towards quite different environments for these dust cells, even though they reside at the same galactocentric distance. 

Most interesting are the prominent spikes in the radial distribution. They point towards radii with an enhanced contribution of unevolved stellar populations to the absorbed energy (see also the circles in the bottom right panel of Fig.~\ref{fig:M31FaceOn}). Two moderate spikes are visible at roughly 6 and 8 kpc. They can be associated with the filamentary structures in the inner disk of M31. A broad peak occurs between 10 and 15 kpc which corresponds to the dust cells in the main star-forming ring. Just outside the 15~kpc radius, a second sharp peak occurs, which is linked to the second large ring of Andromeda. Beyond this radius, there is an overall increase in $\mathcal{F}_\mathrm{unev.}^\mathrm{abs}$, but without any distinct features. At these radii, the overall flux from the galaxy becomes very faint, and the dust grid contains fewer cells. The Monte Carlo noise starts to  dominate, which makes it too speculative to interpret any structure in the outskirts.

It is remarkable that the dust heating fraction by unevolved stellar populations stays so low up to 10 kpc (with only a small number of dust cells as an exception). The radiation field is strong in the centre, where relatively little dust resides. Photons from evolved stellar populations may thus travel through the galaxy without being absorbed. Once they reach the dusty ring at 10 kpc, their chance of being absorbed increases strongly. This explains why the dust heating in the star-forming rings of M31 still has an important contribution from the evolved stellar populations. This effect of non-local heating is a crucial element in energy balance studies and difficult to capture in 2D models of galaxies, or models without geometrical assumptions like energy balance SED fitting.

As a caveat to these results, we remind the reader of the residual maps in Fig.~\ref{fig:residuals}. There, the $FUV$ flux is underestimated in the dustiest regions. The effect is not so large for NIR emission. We argued that this could be caused by non-uniform $FUV$ attenuation across the disk (see Sect.~\ref{sec:Residuals}). In the dustiest regions of our model, there is not enough energy coming from unevolved populations. As a result, we underestimate their contribution to the dust heating fraction. Unfortunately, it is difficult to estimate the magnitude of this effect at the present time.

\subsection{Projected dust heating in M31} \label{sec:ProjectedHeating}

We now look at the dust heating projected on the sky. While the intrinsic model and radiative transfer simulations are still in 3D, we use the 2D maps projected on the sky, recreating the observed line-of-sight for each pixel. This exercise is useful to connect the 3D model to 2D observations. 

As in the previous section, we treat the contributions of the `unevolved' components as one single component. This includes absorbed energy from young and ionizing stellar populations, plus the internally absorbed energy from the subgrid implementation of star-forming regions. We define the total dust heating fraction as the fraction of absorbed energy coming from a particular stellar population. For the unevolved populations for example, we denote this as $F_\mathrm{unev.}$. Naturally, $F_\mathrm{evolved}$ and  $F_\mathrm{unev.}$ add up to unity.

It is not straightforward to split these quantities into separate wavelengths because dust heating is non-local in wavelength. Dust emission at a particular wavelength is caused by the total radiation field and not by the stellar radiation at a single wavelength. Moreover, it depends in a non-linear way on the total absorbed energy. Therefore, while $F_\mathrm{unev.}+ F_\mathrm{evolved} = 1$ on a global scale, this is not true for individual wavelengths. Yet again it is useful to investigate whether warm dust (emitting in the MIR) is heated differently than cold dust (emitting in the submm). 

We follow the definition of \citet{DeLooze2014} for the wavelength dependent dust heating fraction, which is based on the absorbed energy per resolution element in the RT simulations. An alternative definition, based on the total radiation field, was first proposed by \citet{Popescu2000} and described in detail in \citet{Natale2015}. It is possible that this equivalent method is quantitatively offset with our definition. However, both methods are viable to qualitatively explore the wavelength dependency of the dust heating fraction.

\citet{DeLooze2014} proposed a way to approximate $F_{\lambda,\mathrm{evolved}}$ and $F_{\lambda,\mathrm{unev.}}$, while maintaining energy conservation within a single waveband. This involves running the full radiative transfer simulation with different combinations of stellar populations (see again Fig.~\ref{fig:initialSED}, bottom panel), and recording the FIR/submm flux at every wavelength, $S_\lambda$. \citet{DeLooze2014} then define: 
\begin{equation} \label{eq:FprimeUnev}
F^\prime_{\lambda,\mathrm{unev.}} \equiv \frac{1}{2} \frac{S_{\lambda,\mathrm{unev.}}+(S_{\lambda,\mathrm{total}}-S_{\lambda,\mathrm{evolved}})}{S_{\lambda,\mathrm{total}}}
\end{equation}
and
\begin{equation}
F^\prime_{\lambda,\mathrm{evolved}} \equiv \frac{1}{2} \frac{S_{\lambda,\mathrm{evolved}}+(S_{\lambda,\mathrm{total}}-S_{\lambda,\mathrm{unev.}})}{S_{\lambda,\mathrm{total}}}\text{,}
\end{equation}
where $S_{\lambda,\mathrm{tot}}$ is the dust emission heated by evolved, young and ionizing stellar populations.  $S_{\lambda,\mathrm{evolved}}$ is the dust emission when only the evolved stellar components are included in the simulations, and $S_{\lambda,\mathrm{unev.}}$ when only the young and ionizing components are included. Note that there is no need to consider any missing subgrid contribution from the star-forming clouds here. Since we are dealing with emission, this is already included in the total SED of the ionizing stellar component (see Sect.~\ref{sec:NewComp}). In the above definitions, $F^\prime_{\lambda,\mathrm{unev.}}$ and $F^\prime_{\lambda,\mathrm{evolved}}$ are the approximate heating fractions by unevolved and evolved stellar populations. They now add up to unity by definition.

\begin{figure}
	\centering
   	\includegraphics[width=0.45\textwidth]{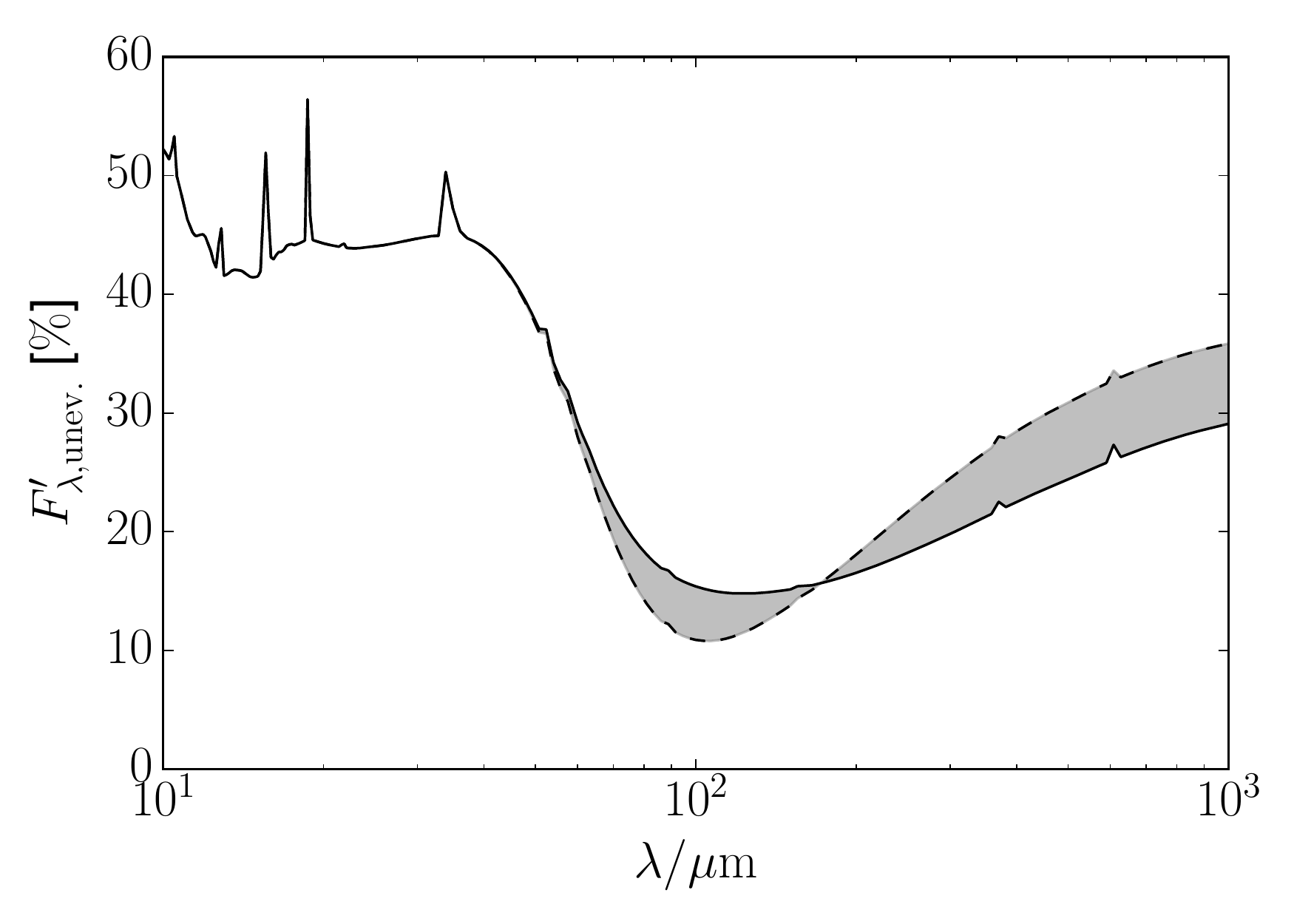}
	\caption{Dust heating fraction as a function of wavelength for the unevolved stellar populations. The solid black line corresponds to the definition in Eq.~\eqref{eq:FprimeUnev}, the dashed line to Eq.~\eqref{eq:FUnev}. The shaded area between the curves indicates the magnitude of the effect of non-local heating in wavelength space.}
	\label{fig:heating}
\end{figure}

This should be compared with the case where the dust heating fractions are not adjusted, i.e.
\begin{equation} \label{eq:FUnev}
F_{\lambda,\mathrm{unev.}} = \frac{S_{\lambda,\mathrm{unev.}}}{S_{\lambda,\mathrm{unev.}}+S_{\lambda,\mathrm{evolved}} }
\end{equation}
and
\begin{equation}
F_{\lambda,\mathrm{evolved}} = \frac{S_{\lambda,\mathrm{evolved}}}{S_{\lambda,\mathrm{unev.}}+S_{\lambda,\mathrm{evolved}} } \text{.}
\end{equation}
We make this comparison to estimate the uncertainty on the dust heating fraction, when looking on a per wavelength basis. Both quantities are shown for the unevolved stellar populations in Fig.~\ref{fig:heating}. The amount of absorbed energy defines the shape and strength of the dust emission spectrum. The area between the curves indicates that the effect of this non-local, non-linear dust heating is most prominent in the submm regime. It can cause a discrepancy of over $5 \%$ in the dust heating fraction. Interestingly, the effect is negligible at shorter wavelengths (< 50 $\upmu$m).

The general shape of the $F^\prime_{\lambda,\mathrm{unev.}}$ curves is qualitatively the same, however. At almost all wavelengths the unevolved stellar populations are \textit{not} the main heating source in M31. From 10 - 50~$\upmu$m, $F^\prime_{\lambda,\mathrm{unev.}}$ is mostly constant at $40 - 50 \%$, with several peaks. These peaks correspond to stochastially heated small grains, sensitive to the high-energy photons coming from the unevolved populations. Beyond 50~$\upmu$m, the contribution of the unevolved populations drops significantly, with a minimal contribution of ~$15\%$ at around 100~$\upmu$m. Interestingly, $F^\prime_{\lambda,\mathrm{unev.}}$ rises again when it comes to heating the colder dust at submm wavelengths. This does not conform to the classical view of dust heating, where unevolved stellar populations have their peak contribution at short wavelengths (10--100~$\upmu$m), but fall off after that, leaving the evolved stellar populations to heat the diffuse dust that emits at submm wavelengths \citep[see e.g.][]{DeLooze2014, Natale2015, Bendo2015}. 

It seems counter-intuitive that the contribution of the unevolved stellar populations starts to increase again beyond 100~$\upmu$m. A possible explanation is that the centre of the galaxy is heated by evolved stars and is warmer (or has a more intense ISRF) than the outer parts of the galaxy. In \citetalias{Smith2012} and \citet{Draine2014} the radiation field in the center of M31 was found strong enough to heat the diffuse dust there to temperatures above $30$ K, so dust emission peaks at shorter wavelengths. Consequently, the emission from the centre of the galaxy contributes more to the global $100\, \upmu$m emission than to the global $\gtrsim 250\, \upmu$m emission.  Hence, when integrating all of the emission over the galaxy, the $100\, \upmu$m emission appears to have a lower $F^\prime_{\lambda,\mathrm{unev.}}$ in Fig.~\ref{fig:heating} than dust at $\gtrsim 250\, \upmu$m.

\begin{figure*}
	\centering
   	\includegraphics[width=1.0\textwidth]{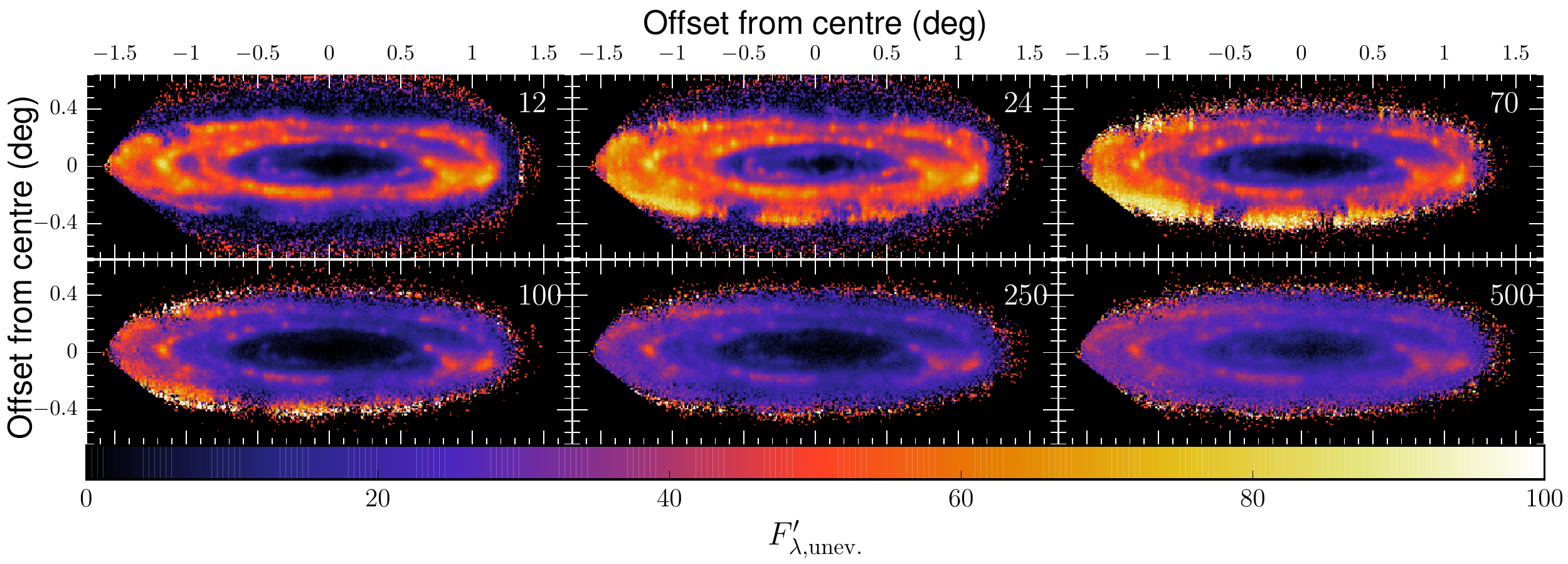}
	\caption{Maps of the dust heating fraction by unevolved stellar populations $F^\prime_{\lambda,\mathrm{unev.}}$ for a selection of wavebands. Top row (left to right): 12, 24 and 70~$\upmu$m. Bottom row (left to right): 100, 250 and 500~$\upmu$m. }
	\label{fig:heatingMaps}
\end{figure*}

When we look at spatial heating maps at different wavelengths, shown in Fig.~\ref{fig:heatingMaps}, we find the classical morphological segregation between disk and bulge, with the influence of the bulge almost stretching out to the 10 kpc ring. In the rings the unevolved stellar populations are dominant at short wavelengths (up to $90\%$ at 24~$\upmu$m). Their contribution slowly declines at longer wavelengths. In the large bulge of M31, the contribution of evolved stellar populations is always high, but peaks at 100~$\upmu$m. There are large spatial variations in the heating fraction at shorter wavelengths. In the submm, these variations are almost gone and the heating fraction is more uniform across the disk.

Going back to the global, aggregate heating fractions (not the approximate, wavelength-dependent quantities), we find that $F_\mathrm{unev.} = 21\%$. In other words, almost $80 \%$ of the total dust emission in Andromeda is due to heating by the evolved stellar populations. This is somewhat lower than the $91 \%$ found in the previous section. This underlines an important difference between the 3D and 2D determination of the dust heating sources. In the projected view, it is possible that the full magnitude of the bulge radiation field is not captured adequately. This results in a lower contribution of the evolved stellar populations to the dust heating. Nevertheless, the evolved stellar populations dominate the global dust heating in both scenarios. Our findings are broadly in line with the dust heating analysis based on Planck colours \citep{PlanckM31}. These authors report that M31 cold dust is mainly heated by the evolved stellar populations as traced by the IRAC 3.6~$\upmu$m emission. They do find that their 70/100~$\upmu$m colour correlates better with the unevolved stellar populations (traced by the 24~$\upmu$m emission). While we do see a rise in $F^\prime_{\lambda,\mathrm{unev.}}$ at wavelengths <100~$\upmu$m, we do not find the unevolved stellar populations to become the dominant energy source. We come back to the point of FIR colours as dust heating tracers in the next section.

\subsection{Tracers of dust heating} \label{sec:heatingTracers}

The analysis we have made so far, involving a full, self-consistent radiative transfer model, is a suitable way to explore the details of dust heating. On the other side, it is very demanding -- both in terms of time and data\footnote{For 512 wavelengths and $2\times10^7$ photon packages per wavelength, running one model takes about $4000$~CPU hours and $15$~GB of RAM.} -- to construct a full RT model for each galaxy one would like to study. Hence, we have looked for possible correlations, in M31, between the total heating fraction and some other observed properties. Here we use the global heating fraction per pixel (i.e. derived by integrating the dust SED) $F_\mathrm{unev.}$. For this analysis, only pixels within a galactocentric radius of $18$ kpc were used. This is to avoid low S/N areas to obscure the main trends. For each correlation, we compute Kendall's rank coefficient $\tau$ \citep{Kendall1938, Kendall1990}. This is a non-parametric correlation coefficient where $\tau = \pm 1$ denotes a perfect (anti-)correlation, and $\tau = 0$ means there is no correlation.

\begin{figure*}
	\centering
   	\includegraphics[width=1.0\textwidth]{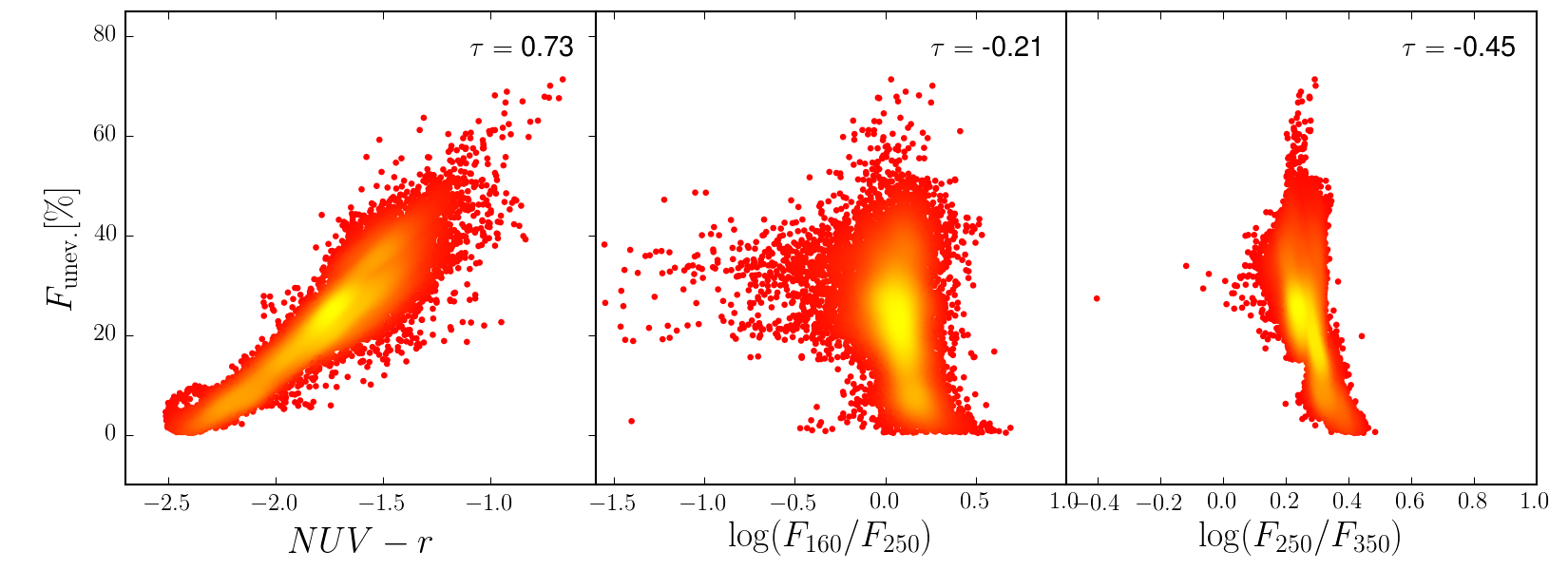}
	\caption{Density plots of the dust heating fraction from unevolved stellar populations $F_\mathrm{unev.}$ vs $NUV-r$ and two commonly used FIR colours. Only pixels within a galactocentric radius of $18$ kpc are shown. Kendall's correlation coefficient $\tau$ is given in each panel. A red colour indicates a small number of data points in that area of the plot, yellow at a large number.}
	\label{fig:heatingColours}
\end{figure*}

In Fig.~\ref{fig:heatingColours}, we plot $F_\mathrm{unev.}$ against $NUV-r$ colour, commonly used in the context of galaxy evolution as an observable tracer of the specific SFR (see e.g. \citealt{daCunha2010,Cortese2012}; \citetalias{Viaene2014}; \citealt{Viaene2016}). There is a clear positive trend ($\tau = 0.73$), where `blue' pixels correspond to higher contributions of unevolved stellar populations to the dust heating than `red' pixels. 

These results are in line with the recent insights by \citet{Boquien2016}, based on panchromatic energy balance SED fitting of sub-galactic regions. They find that sSFR directly reflects the relative contributions to the total radiation field. Vice versa, they find that FIR bands like $70 \, \upmu$m and $100 \, \upmu$m are more sensitive to the absolute radiation field. This is because, FIR bands are much more influenced by dust temperature (so total energy), and less by the relative contributions to the total energy.

We therefore do not observe convincing trends between $F_\mathrm{unev.}$ and two commonly used FIR colours to trace dust heating: the 160/250 and 250/350 flux ratios (middle and right panel of Fig.~\ref{fig:heatingColours}). The dust heating can take any fraction between 0 and $50 \%$ for most values of 160/250 or 250/350 colour. In the former, there is a significant scatter towards red 160/250 colours (or low dust temperatures). These pixels are located in the outer regions of the disk.  They have cold dust, but the influence of unevolved stellar populations is relatively higher than at shorter galactocentric radii.
The 250/350 colour range is particularly narrow, which makes it difficult to use as a predictor.

\begin{figure*}
	\centering
   	\includegraphics[width=1.0\textwidth]{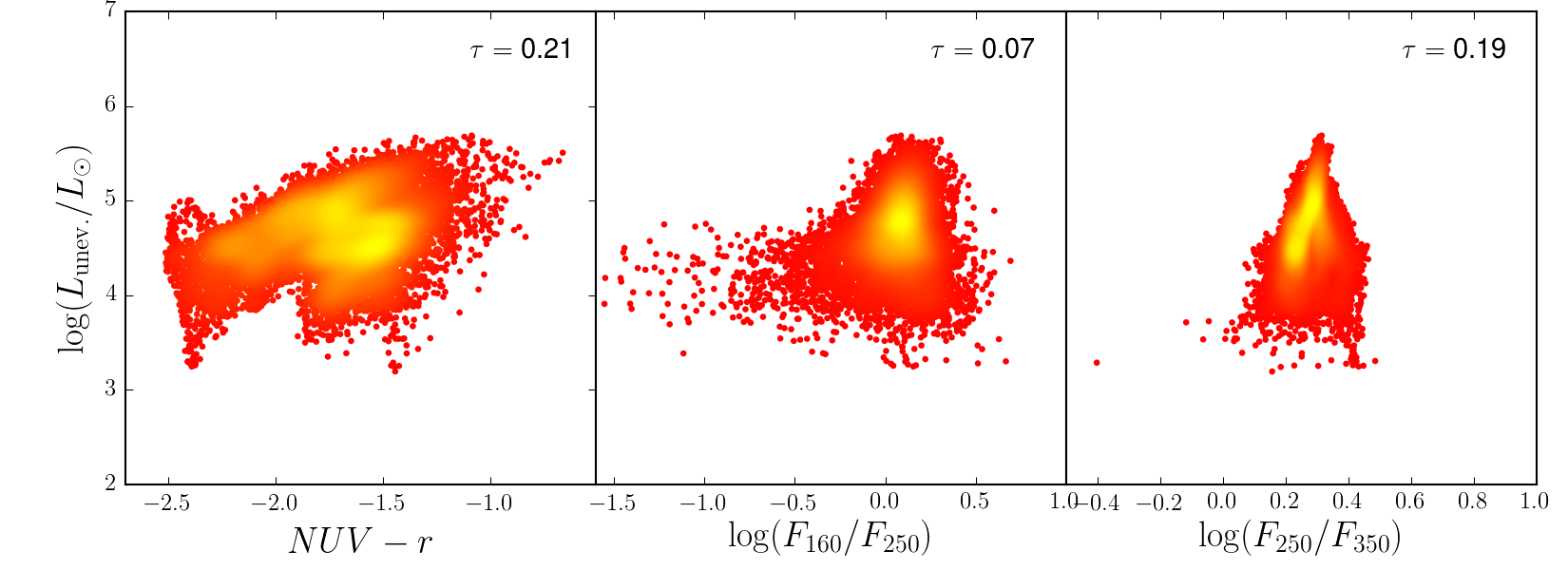}
	\caption{Density plots of the projected dust absorption luminosity from unevolved stellar populations $L_\mathrm{unev.}$ vs $NUV-r$ and two commonly used FIR colours. Only pixels within a galactocentric radius of $18$ kpc are shown. Kendall's correlation coefficient $\tau$ is given in each panel. A red colour indicates a small number of data points in that area of the plot, yellow at a large number.}
	\label{fig:AbsheatingColours}
\end{figure*}

\citet{Bendo2015} also found that the \textit{fraction} of dust heating is hard to predict using FIR colours. On the other hand, they were able to link FIR colours to the \textit{absolute} energy output from evolved and unevolved stellar populations, traced by $3.6 \, \upmu$m and $H\alpha$ emission, respectively. The same results were confirmed for M31 by \citet{PlanckM31}. In our model framework, we can push this one step further and look at the total absorbed energy per pixel for each stellar population. In Fig.~\ref{fig:AbsheatingColours}, we plot this value for the unevolved stellar populations ($L_\mathrm{unev.}$). This quantity is similar to the $L^\mathrm{abs}$ parameters used in Sect.~\ref{sec:3Dheating}, but now projected to the 2D pixels on the sky. 

However, the trends we observe for $L_\mathrm{unev.}$ are less convincing then for $F_\mathrm{unev.}$. $NUV-r$ is still the strongest correlation ($\tau = 0.21$), but still weak and clearly bimodal. For the 250/350 colour, the trend is weak ($\tau = 0.19$) and the colour range is narrow. Again, the 160/250 colour ($\tau = 0.07$) has a large variance and is also not suitable as a predictor for $L_\mathrm{unev.}$. This is rather surprising because in M31, the 160/250 flux ratio should be most sensitive to the shape of the cold dust emission spectrum (see for example Fig.~\ref{fig:initialSED}). The peak of the dust distribution lies at $160 \, \upmu$m, while $250 \, \upmu$m is part of the Rayleigh-Jeans tail of the cold dust emission. One would naively expect a link to the total absorbed energy. It appears that the variation in dust temperature on a local, pixel-by-pixel scale is too large to capture in a single FIR colour.

\begin{figure*}
	\centering
   	\includegraphics[width=1.0\textwidth]{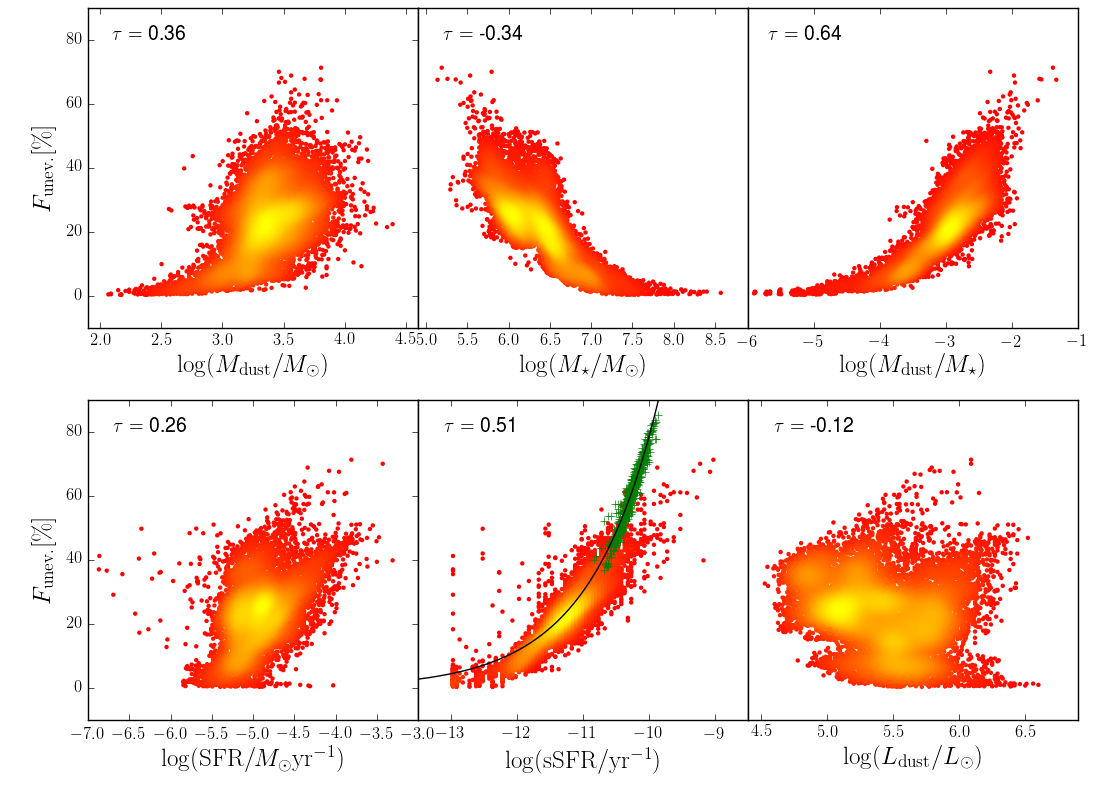}
	\caption{Density plots of the dust heating fraction from unevolved stellar populations $F_\mathrm{unev.}$ vs a number of physical parameters, obtained from \citetalias{Viaene2014}. A red colour indicates a small number of data points in that area of the plot, yellow at a large number. Kendall's correlation coefficient $\tau$ is given in each panel. First row: dust mass, stellar mass, dust-to-stellar mass ratio. Second row: Star formation rate, specific SFR, temperature of the cold dust in the ISM.  The black line is a best fit linear model, the green line is the relation found by \citet{DeLooze2014} for M51. Parameters are derived according to the MAGPHYS SED model.}
	\label{fig:heatingParams}
\end{figure*}

We now look for possible correlations between the dust heating and physical properties. Our previous investigation of M31 \citepalias{Viaene2014} offers a range of physical parameters for each pixel. In Fig.~\ref{fig:heatingParams}, the most commonly used physical parameters are plotted against $F_\mathrm{unev.}$. There is a highly scattered positive trend with $M_\mathrm{dust}$ ($\tau = 0.36$). This could be interpreted in the context of star formation, where dusty regions are often associated with star-forming regions. Consequently, these regions are dominated by the radiation field of unevolved stellar populations. This is in accordance with a positive trend with SFR ($\tau = 0.26$). However, the scatter in this correlation is large, suggesting other elements are in play as well. Interestingly, no clear trend with the total dust luminosity, $L_\mathrm{dust}$, was found either ($\tau = -0.12$). Infrared luminous galaxies often point at high levels of star formation, and so one would naively expect a higher contribution of unevolved stellar populations to the heating fraction. However, for the individual pixel-regions in M31, this is clearly not the case.

We find a negative correlation ($\tau = -0.34$) with stellar mass ($M_\star$). If one assumes that the bulk of stellar mass is in evolved stars, then it is not surprising that the pixels with high stellar mass correspond to low $F_\mathrm{unev.}$. But here as well, the trend is ambiguous and suggests the influence of a third parameter. We further investigate this by looking at the specific dust mass ($M_\mathrm{dust}/M_\star$). This quantity gives an idea of the dust content, relative to the stellar mass. High $M_\mathrm{dust}/M_\star$ values point at dusty regions and those can indeed be associated with high $F_\mathrm{unev.}$ ($\tau = 0.64$). The heating fraction of unevolved stellar populations approaches zero for the lowest specific dust masses. This suggests that at low dust content, there is also no significant radiation field from young or ionizing stellar populations.

Another significant correlation is with sSFR ($\tau = 0.51$). This correlation is likely to be intrinsic as the trends with SFR or $M_\star$ separately show much more scatter. This trend is not completely unexpected, since we already found a positive link between $F_\mathrm{unev.}$ and $NUV-r$ colour, and this colour correlates with sSFR in M31 \citepalias{Viaene2014}. sSFR traces the hardness of the UV radiation field \citep{Ciesla2014} and is linked to several FIR/submm colours in galaxies \citep[e.g. ][]{Boselli2010,Boselli2012,Boquien2011, Cortese2014}. Moreover, \citet{DeLooze2014} found a tight link between the dust heating fraction of unevolved stellar populations and sSFR for M51. We overplot their data points and best fit curve in the central, bottom-row panel of Fig.~\ref{fig:heatingParams}. In M51, higher fractions of $F_\mathrm{unev.}$ can be achieved compared to Andromeda. The regions of M31 follow the average relation for the M51 regions closely for all but the lower sSFR values. In fact, both datasets seem to lie in the continuation of one another. This is quite intriguing, since the physical size along the major axis of the M31 pixels is about $3.5$ times smaller than for the M51 pixels. On top of that is the difference in inclination of both galaxies, where a far greater distance is probed along the line of sight in M31.

The bottom line is that extensive galaxy parameters such as $M_\mathrm{dust}$, $L_\mathrm{dust}$, $M_\mathrm{star}$ or SFR are not suitable to disentangle the dominant dust heating sources. A better option is to resort to intensive parameters that somehow express the ratio of evolved and unevolved stellar populations, e.g. $M_\mathrm{dust}/M_\star$ or sSFR. While there is still some variance in the correlation with sSFR, it is small enough to identify the dominant heating source. Of course, a more thorough investigation involving multiple galaxies is desirable to quantify this relation.

\section{Conclusions} \label{sec:conclusions}

We have constructed a highly detailed model for the Andromeda galaxy to investigate the dust heating mechanisms in this galaxy. Our model is based on observed morphologies and uses 3D panchromatic radiative transfer to simulate the dust-starlight interactions in a realistic setting. The main points of this work are:
\begin{itemize}
\item The integrated SED of M31 is fitted well and the resulting attenuation curve is consistent with observations, but with a broader UV bump. We are able to constrain 2 of our 3 free parameters: the dust mass and the luminosity of the young stellar component. The intrinsic luminosity of the ionizing stellar population is more difficult to constrain.

\item The model is able to reproduce the observed morphologies fairly well from far-UV to submm wavelengths. The median (absolute value) deviation between model and observations across all bands is $22 \%$. The flux in the rings is generally underestimated, and the flux in the inter-ring regions is overestimated. Lowering the vertical scale height of the galaxy somewhat mitigates this effect, but cannot resolve it. The discrepancies are a combination of deprojection effects, variations in the nature and size distribution of the dust grains, and the subgrid treatment of the star-forming regions. With an inclination of $77.5 \degree$, M31 is about the limiting case for these kind of deprojected radiative transfer models.

\item The dust in Andromeda is mainly heated by the evolved stellar populations. From a 3D analysis of the radiation field, we find that $91 \%$ of absorbed stellar radiation originates in evolved stellar populations. This high value is mainly due to the bright bulge, which dominates the radiation field out to the main star-forming ring at 10 kpc. Inside and beyond the star-forming ring, the contribution of unevolved stellar populations (i.e. ionizing and non-ionizing young stellar populations) to the radiation field increases, but usually remains in the $10-30 \%$ range.

\item The sSFR (and its observational counterpart, $NUV-r$ colour) is a promising tracer for the total dust heating fraction and thus for the relative contributions to the total IR emission. We find that regions in M31 match the best fit relation derived from M51 pixels. In fact, the two datasets make a rather smooth and continuous sequence. More research is required to assess whether sSFR (and $NUV-r$) is a general tracer of dust heating fractions in galaxies at a local scale.

\end{itemize}

Our study has shown that the heating of dust by stellar populations is a complex problem with a large influence of geometry in three dimensions. Effects like non-local heating make it difficult to draw conclusions from 2D `on-sky' analysis. The contribution of evolved stellar populations is dominant in M31, and can also be significant in other galaxies. One should therefore be careful to directly link dust emission to the properties of unevolved stellar populations.

As a final remark, we want to underline that this is one of the first attempts to construct a detailed geometrical model of stars and dust of a resolved and moderately inclined galaxy, based purely on 3D radiative transfer simulations. Because 3D radiative transfer models are computationally demanding, we were forced to make a number of simplifying assumptions on the geometry, and limit the number of free parameters in the model. In the future, we will refine our method on several points and apply it to galaxies of different morphologies and inclination.

\begin{acknowledgements}

This project has received funding from DustPedia, a European Union’s Seventh Framework Programme for research, technological development and demonstration under grant agreement no 606874 \\ 
S.V gratefully acknowledges the support of the Flemish Fund for Scientific Research (FWO-Vlaanderen). IDL is a post-doctoral researcher of the FWO-Vlaanderen (Belgium). ET and AT acknowledge the support from the ESF grants IUT40-2, IUT26-2. \\
M. Boquien acknowledges funding by the FIC-R Fund, allocated to the project 30321072. \\
This work has been realized in the frame of the CHARM framework (Contemporary physical challenges in Heliospheric and AstRophysical Models), a phase VII Interuniversity Attraction Pole (IAP) programme organised by BELSPO, the BELgian federal Science Policy Office. \\
We thank all the people involved in the construction and the launch of \textit{Herschel}. SPIRE has been developed by a consortium of institutes led by Cardiff University (UK) and including Univ. Lethbridge (Canada); NAOC (China); CEA, LAM (France); IFSI, Univ. Padua (Italy); IAC (Spain); Stockholm Observatory (Sweden); Imperial College London, RAL, UCL-MSSL, UKATC, Univ. Sussex (UK); and Caltech, JPL, NHSC, Univ. Colorado (USA). This development has been supported by national funding agencies: CSA (Canada); NAOC (China); CEA, CNES, CNRS (France); ASI (Italy); MCINN (Spain); SNSB (Sweden); STFC and UKSA (UK); and NASA (USA). HIPE is a joint development (are joint developments) by the \textit{Herschel} Science Ground Segment Consortium, consisting of ESA, the NASA \textit{Herschel} Science Center, and the HIFI, PACS and SPIRE consortia.\\
\end{acknowledgements}

\bibliographystyle{aa} 
\bibliography{allreferences}

\newpage

\appendix

\section{Model components} \label{subsec:modcomponents}

For our aim to investigate dust heating properties, we simplify the galaxy to a handful of `average' galaxy components. They are listed in Table~\ref{tab:components} and we discuss them in more detail below.

\subsection{Evolved stellar populations} \label{subsec:OldGeometry}
The 3D structure of the evolved stellar populations (> 100 Myr~\footnote{In this study, we consider stellar populations to be evolved when their UV radiation is not significant any more and they mainly produce optical/NIR emission.}) is a necessary input for our model. We simplify this distribution to two components: a disk and a bulge. For M31, we derive the distribution of the evolved stellar populations from the IRAC 3.6~$\upmu$m image. The derived geometries must be deprojected to construct their 3D equivalent. Deprojecting an image assumes all features reside in the same (flat) disk. The bulge, however, has a large vertical dimension, so it would be smeared out into a bar-like structure in the deprojected image. This issue is avoided by subtracting the bulge from the IRAC 3.6~$\upmu$m image before deprojection. An accurate representation of the bulge of M31 is needed in 2D - to subtract from the 3.6~$\upmu$m image - but also in 3D, to implement it in our RT model. We thus require an analytical model that can be fitted to the 2D image, but can be deprojected in a realistic way. The big issue here is the boxiness of Andromeda's bulge \citep{Beaton2007}. 

Several analytical bulge-disk decompositions are available for M31 \citep[e.g.,][]{Courteau2011, Tamm2012}, but they all produce elliptical isophotes. Using N-body simulations, \citet{Athanassoula2006} managed to construct a 3D model that approximately reproduced the boxy bulge of M31. They argue that a bar is responsible for the observed boxiness. Unfortunately, they do not provide an analytical prescription. The fitting of 2D analytical boxy bulges is possible using e.g. GALFIT \citep{Peng2010}. However, to our knowledge there is no 3D analytical prescription to mimic a boxy bulge. Deriving such a prescription falls beyond the scope of this paper. We therefore resort to the next best 3D representation of Andromeda's bulge: the flattened Einasto profile derived by \citet{Tamm2012} (see their equation B.4). It describes the surface brightness $S(r)$ as a function of radius $r$:
\begin{equation}
S(r) = S_{\text{c}}\, \exp\left\{-d_N \left[ \left( \frac{r}{a_{\text{c}}}\right)^{1/N}-1\right]
    \right\}.
\end{equation}
This prescription has two free parameters: the concentration radius of $a_c = 1.155$ kpc which contains half of the total brightness, and a structure parameter $N = 2.7$. The numerical constant $d_N$ is only dependent on $N$ and ensures that half of the brightness is containted in $a_c$. An intrinsic flattening in the vertical direction of $q = 0.72$ was imposed in the sense that
\begin{equation}
S(x,y,z) = \frac{1}{q}\, S_{\text{c}}\left(\sqrt{x^2 + y^2 + \frac{z^2}{q^2}}\right).
\end{equation}

We find a difference in position angle of $13.8 \degree$ between the major axis of the bulge and the major axis of the disk. The flattened \citet{Sersic1963} bulge derived by \citet{Courteau2011} was also tested, but the resulting disk (subtracting the bulge from the IRAC 3.6~$\upmu$m image) showed stronger artefacts.

\citet{Tamm2012} found a bulge-to-disk ratio ($B/D$) of 0.54 from SDSS-$i$ band imaging and we find that this ratio also works well at 3.6~$\upmu$m. Upon subtraction of the bulge model from the 3.6~$\upmu$m image, some pixels show a slight over- or under-subtraction caused by the small deviation of the bulge in M31 from the smooth Einasto profile. To ensure no negative emission in the centre of the disk, we replace the inner circle with a radius of 5 pixels (685 pc) by the local disk background just like we have masked M32 (see Sect.~\ref{sec:data}). This correction is unlikely to affect the radiation field in the nucleus of M31 as the bulge is dominant here.

The corrected, bulge-subtracted 3.6~$\upmu$m image was used as input geometry for the evolved stellar disk. Following \citet{DeLooze2014}, we assume an exponential vertical profile for the disk. In practice, this introduces an extra parameter: the vertical scale height. We rely on the modelling efforts of \citet{Tempel2010}, who derive radial scale lengths and flattening parameters for Einasto profile fits to M31. They take the 3D nature of the galaxy into account by concurrently modelling the extinction by dust. Their results are equivalent to a vertical scale height of 538 pc for the evolved stellar disk. This value also matches the average scale height found by \citet{DeGeyter2014} for a sample of 12 edge-on galaxies.

A panchromatic approach requires each model component to have a luminosity at every UV to submm wavelength. In practice this is achieved by selecting a suitable template SED for each component. The luminosity is then scaled to match a certain value at a representative wavelength. For the evolved stars we make use of the \citet{Bruzual2003} simple stellar populations (SSPs). In this paper, we make use of their low resolution version of the \textit{Padova 1994} \citep{Padova1994} model, which uses the \citet{Chabrier2003} initial mass function. The templates are characterised by metallicity and age.

For the bulge, we assume an average age of 12 Gyr and Solar metallicity ($Z = 0.02$). In the disk, we assume the same metallicity and an average age of 8 Gyr for the stars. The estimates for age and metallicity are taken from \citet{Saglia2010}, who analysed several long-slit spectra across the bulge and inner disk of M31. We normalize this SED for both bulge and disk to the total IRAC 3.6~$\upmu$m luminosity and use $B/D = 0.54$. This is the only stellar SED we can unambiguously normalize as the IRAC 3.6~$\upmu$m band is a "pure" band, i.e. it traces only one component (evolved stars) and suffers least from attenuation by dust.

\subsection{Young stellar populations} \label{sec:YoungComp}

Following the general picture of large galaxies, we assume that the young stellar populations reside in a disk that is thinner than the evolved stellar disk. These stars are bright in the $UV$, but can not ionize hydrogen molecules. They should have had the time to migrate from or dissolve their dusty birth clouds, which is typically after $10^6$-$10^7$ yr \citep{Hartmann2001,Murray2011,Bailey2014}. Their emission is therefore only obscured by dust in the diffuse ISM. 

Evolved stars can contribute significantly to the FUV flux \citep[e.g.][]{Kennicutt2009}. The bulge of M31 is known to have old stars which contribute to the UV flux \citep{Saglia2010, Rosenfield2012}. We correct for this using the formula derived by \citet{Leroy2008} to obtain $S_{FUV}^\mathrm{young}$, the surface brightness from young stellar populations in the $FUV$:
\begin{equation} \label{eq:fyoung}
S_{FUV}^\mathrm{young} = S_{FUV} - \alpha_{FUV} S_{\mathrm{IRAC}\, 3.6}\text{.}
\end{equation}
In \citetalias{Ford2013} this was applied to M31 and $\alpha_{FUV} = 8.0 \times 10^{-4}$ was found, based on the same FUV and IRAC 3.6~$\upmu$m observations of M31 as used in this work. Note that this correction only affects the geometrical distribution of the young stellar population. In practice, this distributes most of these stars in the disk, and very little in the bulge.
The corrected $FUV$ image is used as the 2D input geometry for the young stellar populations. Similar to the disk of evolved stellar populations in the previous section, we add an exponential profile in the third dimension. We set the vertical scale height to be 190 pc, following \citet{Tempel2010}.

We assume that the relative surface brightness in the FUV band gives a good approximation of the distribution of these stars, even though the intrinsic flux may differ from the observed one. This is a first order approximation as dust extinction is not uniform across the disk. Regions with visible dust lanes are likely to absorb a larger fraction of the $FUV$ radiation from young stars. This will decrease their relative importance in the map. Vice versa, the distribution of young stars will be boosted in the inter-ring regions.

Ideally, one would independently vary the amplitude of the FUV emissivity (for the fixed assumed vertical distribution) for each resolution element on the images. It is clear that such an approach is currently intractable, as it would require far too many RT calculations to explore all permutations of the FUV brightness distribution.

As SED template for this component, a \citet{Bruzual2003} template was again used. There is a known metallicity gradient in M31 \citep[e.g., ][HELGA V]{Mattsson2014}, but it is shallow within the optical disk. As we are working with average stellar components, we keep the metallicity at the Solar value. We set a mean age of 100 Myr, following \citet{DeLooze2014}. Using a different age would mainly affect the UV flux of this component, and change the normalization of the SED. However, we found that this has only little effect on the FIR/submm SED. The SED is normalized to an intrinsic FUV luminosity.

The total intrinsic $FUV$ in M31 can of course not be measured directly, so we leave this parameter free. We can, however, make an educated guess. Using Eq.~\eqref{eq:fyoung} we find that the contribution of young stellar populations to the total FUV flux is about $84\%$. The extinction in the FUV band of the global SED model of \citetalias{Viaene2014} is $45 \%$ in flux. When we correct the observed FUV flux with these two factors, we obtain an initial guess of the intrinsic FUV flux for the young stellar component.

\subsection{Ionizing stellar populations} \label{sec:NewComp}

Embedded star formation is a crucial element in the energy balance between UV/optical (absorption) and FIR/submm (emission). This was first recognized by \citet{Silva1998} and \citet{Popescu2000}, and confirmed by consequent work \citep{Misiriotis2001, Baes2010, Popescu2011, MacLachlan2011,DeLooze2012a,DeLooze2012b,DeLooze2014}. The ionizing stars (10 Myr) are still embedded in their birth clouds and produce a hard, ionizing radiation. Due to the dust in these clouds it is near impossible to trace their emission directly. Ionizing stars ionize their surrounding gas. The ionized \ion{H}{II} gas can be traced through its H$\alpha$ emission. \citet{Devereux1994} mapped a large part of Andromeda in H$\alpha$. Unfortunately their map does not cover the entire disk and there are several artefacts present after continuum subtraction. Another indirect tracer is the warm dust surrounding the stellar birth clouds. The emission of warm dust was already derived for M31 in \citetalias{Ford2013}, using the MIPS 24~$\upmu$m image. This waveband can hold a significant contribution of warm dust residing in the atmospheres of evolved stars, \citet{Leroy2008} therefore derive the following correction to obtain $S_{FUV}^\mathrm{ion.}$, the surface brightness of the ionizing stellar component in the $FUV$:
\begin{equation} 
S_{FUV}^\mathrm{ion.} = S_{24} - \alpha_{24} \times S_{\mathrm{IRAC}\, 3.6}\text{.}
\end{equation}
This was also applied to M31 in \citetalias{Ford2013}. They found that $\alpha_{24} = 0.1$. Despite removing this contribution, the map may still be contaminated by emission from diffuse dust, heated by the general ISRF \citep[see also][]{Kennicutt2009}. One should keep in mind that this seriously hampers the use of the 24~$\upmu$m band as quantitative tracer of the total SFR. However, as we are only interested in the spatial distribution of the obscured SFR, the effect of diffuse dust emission is less important.
In the third (vertical) dimension, we again use an exponential profile, as for the young and evolved stellar disks. The vertical scale height is set to 190 pc, similar as for the young stellar disk.

For the SED of the ionizing stars we follow the approach of \citet{DeLooze2014} and use the standard MAPPINGS III template SEDs for obscured star formation from \citet{Groves2008}. These SEDs are isotropic representations of the \ion{H}{II} regions around massive clusters of young stars. We adopt the same values for metallicity ($Z = 0.02$), compactness ($\log \mathcal{C}=6$), pressure of the surrounding ISM ($P_0 = 10^{12}$ K/m$^3$) and cloud covering factor ($f=0.2$) as in \citet{DeLooze2014} and refer the reader to this work for more specifics about the use of these templates in SKIRT. 

As for the young stellar component, we also normalize this SED to the total intrinsic FUV emission of the ionizing stars. For the same reason (extinction by dust) we leave this intrinsic luminosity as a free parameter. We can again make an initial guess for this parameter: the \citet{Groves2008} templates are normalized to a SFR of 1~$M_\odot$~yr$^{-1}$. We take the template luminosity in the FUV band and scale this to 0.2~$M_\odot$~yr$^{-1}$, which is a good approximation of the SFR in M31 (\citealt{Tabatabaei2010}; \citetalias{Ford2013, Viaene2014}; \citealt{Rahmani2016}). We use this value as an initial guess for the intrinsic FUV luminosity of the ionizing stars.

\subsection{Interstellar dust} \label{subsec:DustGeometry}

To map the dust in a galaxy one can resort to one of the FIR/submm bands. However, as dust emission peaks at different wavelengths depending on its temperature, choosing a single band is not straightforward. 

Alternatively a map of the UV attenuation A$_{FUV}$ can be created  \citep{Montalto2009,DeLooze2014}. This method takes into account multiple FIR/submm wavebands and links this to the FUV morphology, where dust can be seen in extinction. Consequently, it should produce a tighter constraint on the dust morphology. There are many recipes available for A$_{FUV}$, based on different calibrations \citep{Cortese2008,Hao2011,Boquien2012,Viaene2016}. The drawback of these recipes is that they indirectly rely on simplified geometries to convert optical depth to dust attenuation.

For M31 in particular, a high-resolution dust attenuation map was constructed by \citet{Dalcanton2015} from extinction of individual stars. Unfortunately, their map only covers one third of our field of view. \citet{Draine2014} constructed a dust mass map based on pixel-by-pixel SED fits from MIR to submm wavelengths. Independently, in \citetalias{Viaene2014}, we performed a panchromatic pixel-by-pixel fit of M31. Their dust map is also consistent - within their model - with the attenuation of UV and optical light. They found that their map was consistent with that of \citet{Draine2014}, within the model uncertainties. 

We will use the \citetalias{Viaene2014} dust mass map as an input for the relative dust mass distribution in our model, but do not fix the total dust mass. To this 2D geometry we again add an exponential profile in the third dimension. \citet{Tempel2010} find a dust scale height of 238 pc, which yields a dust-to-stellar scale height $\lesssim 0.5$. This is consistent with RT models of edge-on galaxies \citep{Xilouris1999,Bianchi2007,DeGeyter2014}.

We use the THEMIS dust model compiled by \citet{Jones2013}. The model consists of amorphous hydrocarbons and silicates and we use their parametrization for diffuse dust. THEMIS is a framework of dust properties, based of a vast set of laboratory data, and characterized by a series of free parameters. We adopt the parametrization that best suits Milky Way dust \citep{Jones2013, Ysard2015}. This model naturally produces the spectral shape in the MIR/FIR/submm, and is consistent in explaining both extinction and emission by the same dust mixture for diffuse dust in the Milky Way \citep{Ysard2015}.

\begin{figure*}
	\centering
   	\includegraphics[width=1.0\textwidth]{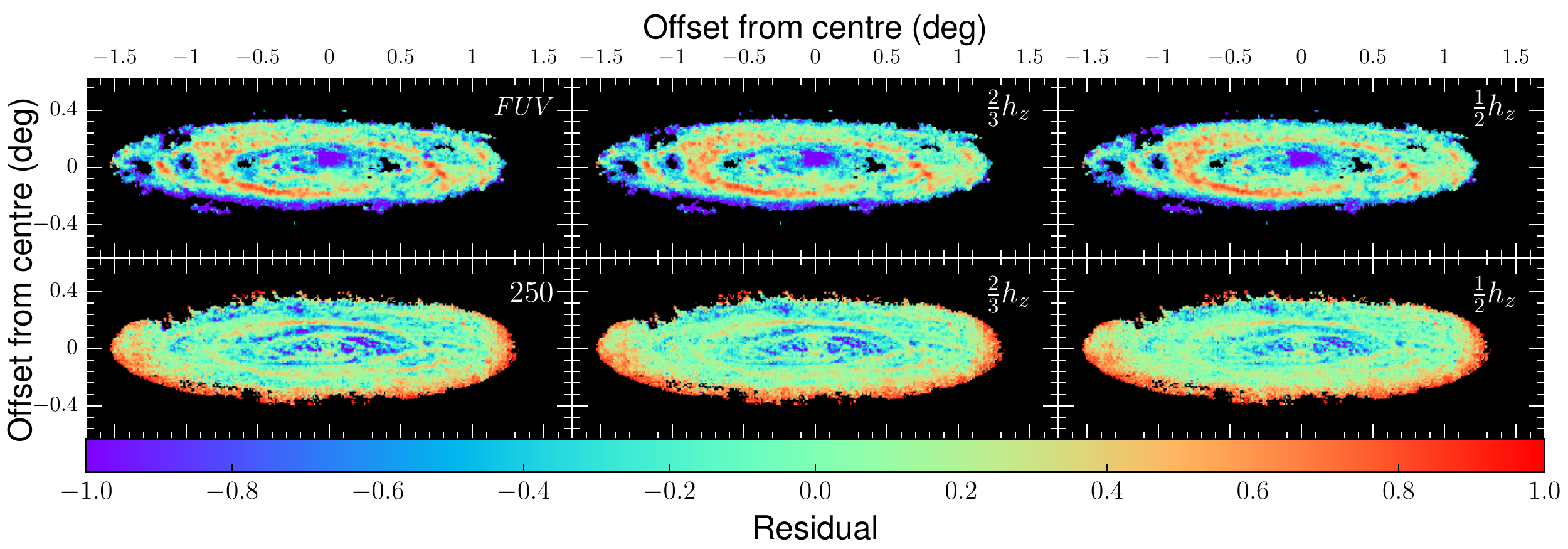}
	\caption{Effect of galaxy thickness on residuals in the $FUV$ band (top row) and at 250~$\upmu$m (bottom row). The first column shows residuals for the original thickness, middle and right columns for 2/3 and 1/2 of that thickness, respectively.}
	\label{fig:ScaleHeightResiduals}
\end{figure*}

The dust component in our RT model is normalized by the total dust mass. The amount of diffuse dust was determined from the dust mass map from \citetalias{Viaene2014}. However, a correction for dust in local star-forming clouds is necessary. Indeed, the SED of the obscured ionizing stars holds a certain amount of dust (see Sect.~\ref{sec:NewComp}). The total observed dust mass in M31 is the sum of this star-forming cloud dust, and the diffuse dust component. This introduces no extra free parameters, but couples the normalization of the total dust mass to the normalization of the SED of the ionizing stars.

We leave the total dust mass as a free parameter. As an initial guess for this value we take the total dust mass given from the MAGPHYS dust map, which is normalized using the \citet{Dunne2000} dust extinction coefficient $\kappa_{350\,\upmu\mathrm{m}} = 0.454$ m$^2$/kg. In the \citet{Jones2013} model, the average $\kappa_{350\,\upmu\mathrm{m}}$ becomes 0.305~m$^2$/kg. This enhances the total dust mass by a factor of $1.5$. However, we note that this average $\kappa_{350\,\upmu\mathrm{m}}$ should in general not be used to convert dust emission into dust masses. The value for the extinction coefficient varies strongly with dust mixture and is still an average factor, with large variance. Still, for our initial guess, it is sufficient.

\section{Observed fluxes} \label{sec:Fluxes}

Table~\ref{tab:Fluxes} lists the integrated fluxes for M31 used in this paper. These are the integrated fluxes from the region we consider in our model. This an ellipse with center at $\alpha =$~00:43:06.28 and $\delta =$~+41:21:12.22, semi-major axis of $1.425 \degree$, semi-minor axis of $0.400 \degree$ and a position angle of $38.1 \degree$. The values may differ slightly from other literature measurements, based on different apertures.

\begin{table} 
\caption{Integrated fluxes for M31 adopted in this paper, listed by increasing central wavelength.}
\label{tab:Fluxes}
\centering     
\begin{tabular}{lll}
\hline
Instrument \& band		& Flux (Jy) & Error (Jy) \\
\hline
\hline
GALEX $FUV$  			& 1.325	& 0.066	\\
GALEX $NUV$ 				& 2.445	& 0.073	\\
SDSS $u$  				& 17.77	& 0.36	\\
SDSS $g$ 				& 80.9	& 1.6	\\
SDSS $r$ 				& 177.2	& 3.5	\\
SDSS $i$ 				& 270.8	& 5.4	\\
SDSS $z$ 				& 328.5	& 6.6	\\
WISE 1  					& 280.8	& 6.7	\\
IRAC 1   				& 280	& 23		\\
IRAC 2   				& 158	& 11		\\
WISE 2  					& 155	& 4.3	\\
IRAC 3   				& 215	& 48		\\
IRAC 4					& 209	& 35		\\
WISE 3					& 203	& 9.1	\\
IRAS $12 \, \upmu$m	 	& 181	& 24		\\
WISE 4 					& 160	& 9.1	\\
MIPS $24 \, \upmu$m   	& 118	& 47		\\
IRAS $25 \, \upmu$m  	& 126	& 17		\\
IRAS $60 \, \upmu$m    	& 820	& 110	\\
MIPS $70 \, \upmu$m    	& 1042	& 104	\\
PACS $100 \, \upmu$m   	& 3389 	& 339	\\
IRAS $100 \, \upmu$m   	& 3570	& 470	\\
PACS $160 \, \upmu$m   	& 7397	& 739	\\
SPIRE $250 \, \upmu$m  	& 5858	& 410	\\
SPIRE $350 \, \upmu$m  	& 3061	& 214	\\
Planck $350 \, \upmu$m 	& 2900	& 300	\\
SPIRE $500 \, \upmu$m  	& 1317	& 92		\\
Planck $550 \, \upmu$m 	& 980	& 100	\\
Planck $850 \, \upmu$m 	& 291	& 14		\\
\hline
\end{tabular}
\end{table}

\section{Alternative scale heights} \label{sec:AltScaleHeights}

The best-fitting model for M31 tends to underestimate the flux in the rings, and overestimate the flux in the inter-ring regions. We test whether this could be induced by the addition of a vertical dimension.

We therefore ran the best-fitting model with smaller scale heights for stars and dust. We show the effect on the $FUV$ and 250~$\upmu$m band in Fig.~\ref{fig:ScaleHeightResiduals}. If the galaxy model is 1/3 thinner, we observe little effect on the residuals. The median deviation from observations decreases from $34\%$ to $30\%$ in the $FUV$ and from $23\%$ to $22\%$ at 250~$\upmu$m. Running the model at only 1/2 of the original thickness slightly decreases these values again. The median deviation from observation now becomes $29\%$ in the $FUV$ and $21\%$ at 250~$\upmu$m.

So, even with a lower scale length, the model still overestimates the flux in the inter-ring regions and underestimates the flux in the rings. We therefore choose not to change our initial assumptions about the scale height of the individual components as these are compatible with the empirical optical depth study of \citet{Tamm2012}.

\end{document}